# Effect of paramagnetic impurities on superconductivity in polyhydrides:

# *s*-wave order parameter in Nd-doped LaH$_{10}$

Dmitrii V. Semenok,[1,*] Ivan A. Troyan,[2] Andrey V. Sadakov,[5] Di Zhou,[1,*] Michele Galasso,[1] Alexander G. Kvashnin,[1] Ivan A. Kruglov,[8,9] Alexey A. Bykov,[3] Konstantin Y. Terent'ev,[4] Alexander V. Cherepahin,[4] Oleg A. Sobolevskiy,[5] Kirill S. Pervakov,[5] Alexey Yu. Seregin,[2,13] Toni Helm,[12] Tobias Förster,[12] Audrey D. Grockowiak,[10,11] Stanley W. Tozer,[11] Yuki Nakamoto,[7] Katsuya Shimizu,[7] Vladimir M. Pudalov,[5,6] Igor S. Lyubutin,[2] and Artem R. Oganov[1,*]

[1] Skolkovo Institute of Science and Technology, 121205, Bolshoy Boulevard 30, bld. 1, Moscow, Russia

[2] Shubnikov Institute of Crystallography, Federal Scientific Research Center Crystallography and Photonics, Russian Academy of Sciences, 59 Leninsky Prospekt, Moscow 119333, Russia

[3] NRC "Kurchatov Institute" PNPI, Gatchina, Russia

[4] Kirensky Institute of Physics, Krasnoyarsk, Russia

[5] P. N. Lebedev Physical Institute, Russian Academy of Sciences, Moscow 119991, Russia

[6] National Research University Higher School of Economics, Moscow 101000, Russia

[7] KYOKUGEN, Graduate School of Engineering Science, Osaka University, Machikaneyamacho 1-3, Toyonaka, Osaka, 560-8531, Japan

[8] Dukhov Research Institute of Automatics (VNIIA), Moscow 127055, Russia

[9] Moscow Institute of Physics and Technology, 9 Institutsky Lane, Dolgoprudny 141700, Russia

[10] Brazilian Synchrotron Light Laboratory (LNLS/Sirius), Brazilian Center for Research in Energy and Materials (CNPEM), Campinas, Brazil

[11] National High Magnetic Field Laboratory, Florida State University, Tallahassee, Florida, 32310, USA

[12] Hochfeld-Magnetlabor Dresden (HLD-EMFL), Helmholtz-Zentrum Dresden-Rossendorf (HZDR), 01328 Dresden, Germany

[13] National Research Center "Kurchatov Institute", Moscow 123182, Russia

**\*  Corresponding authors:** Artem R. Oganov (a.oganov@skoltech.ru), Di Zhou (D.Zhou@skoltech.ru), Dmitrii Semenok (dmitrii.semenok@skoltech.ru)

## Abstract

Polyhydrides are a novel class of superconducting materials with extremely high critical parameters, which is very promising for applications. On the other hand, complete experimental study of the magnetic phase diagram for the best so far known superconductor, lanthanum decahydride LaH$_{10}$, encounters a serious complication because of the large upper critical magnetic field $H_{C2}(0)$, exceeding 120–160 T. Partial replacement of La atoms by magnetic Nd atoms results in a decrease of the upper critical field, which makes it attainable for existing pulse magnets. We found that addition of neodymium leads to significant suppression of superconductivity in LaH$_{10}$: each atomic % of Nd causes decrease in $T_C$ by 10–11 K. Using strong pulsed magnetic fields up to 68 T, we constructed the magnetic phase diagram of the ternary (La,Nd)H$_{10}$ superhydride, which appears to be surprisingly linear with $H_{C2} \propto |T - T_C|$. The pronounced suppression of superconductivity in LaH$_{10}$ by magnetic Nd atoms and the robustness of $T_C$ with respect to nonmagnetic impurities (e.g., Y, Al, C) under Anderson's theorem indicate the isotropic (*s*-wave) character of conventional electron-phonon pairing in the synthesized superhydrides.

# Introduction

The search for high-temperature superconductivity is one of the most challenging tasks of condensed matter physics and materials science.[1] An appealing idea dating back to the early 1960s was that metallic hydrogen should be a high-temperature superconductor with a critical temperature $T_C$ of 300–400 K.[2,3] However, high pressures above 3–4 Mbar are needed for the transition of hydrogen from the molecular insulating phase to the metallic state.[4] For this reason, researchers have recently started to explore the possibility of inducing a superconducting state by combining other elements to hydrogen,[5] which led to the formation of a novel class of chemical compounds — polyhydrides — with a significant reduction of the metallization pressure while maintaining high $T_C$.

Several hydrogen-rich superconductors, such as $H_3S$ ($T_C$ reaches 191–204 K), $YH_6$ ($T_C$ = 224–227 K) and lanthanum superhydride $LaH_{10}$ ($T_C$ is about 250 K), have been successfully synthesized by several groups.[6-10] Because of its extremely high critical temperature, the lanthanum–hydrogen system has become of particular importance and been intensively studied. Key to a better understanding of polyhydrides is an exploration of their magnetic phase diagrams, which show the boundary between the superconducting and normal state of polyhydrides in the "temperature–magnetic field" coordinates, and enable to verify the type of the superconducting state. However, such comprehensive experimental studies of $LaH_{10}$ encounter a great technical challenge due to very high values of the upper critical magnetic field $\mu_0 H_{C2}(0)$ exceeding 120–160 T,[9] whereas modern pulsed magnets can only generate fields of up to 70–100 T in a volume of a high-pressure diamond anvil cell (DAC).

As we show in this work, $\mu_0 H_{C2}(0)$ can be significantly reduced by the partial replacement of the La atoms without disrupting the cubic crystal structure of lanthanum superhydride $LaH_{10}$. This enables investigations in existing pulsed magnets. Moreover, the study of the effect of nonmagnetic (e.g., Y) and magnetic (such as Nd) impurities or dopants on superconductors makes it possible to distinguish between conventional and unconventional mechanisms of superconductivity.[11]

Anderson's theorem[12] states that nonmagnetic (scalar) impurities do not affect the isotropic singlet $s$-wave order parameter in the conventional Bardeen–Cooper–Schrieffer (BCS) theory[13], whereas scattering on paramagnetic centers is very efficient in destroying $s$-wave electron–electron pairing.[11] Nonmagnetic and magnetic impurities are equally detrimental to the critical temperature $T_C$ of unconventional superconducting states.[14] The introduction of such impurities can provide important information on the structure of the magnetic phase diagram and the mechanism of pairing in polyhydrides under pressure.

In this work, we successfully synthesized a series of ternary polyhydrides $(La,Nd)H_{10}$ containing 8–20 at% of Nd at a pressure of 170–180 GPa. Using pulsed magnetic fields up to 68 T, we constructed the magnetic phase diagram of $(La,Nd)H_{10}$, which showed a surprisingly linear $\mu_0 H_{C2}(T) \propto (T_C - T)$. The results of the transport measurements indicate that the lanthanum superhydride follows the conventional Bardeen–Cooper–Schrieffer (BCS) model of superconductivity with the $s$-wave pairing.

# Results

**High-pressure synthesis and stability of La–Nd hydrides**

Several La–Nd alloys with approximately 6.5, 8, 9, 20, 25, and 50 at% of Nd were prepared using melting of La and Nd in an inert atmosphere. The samples were carefully characterized by the scanning electron microscopy and X-ray powder diffraction (Supporting Information Figures S1–S4). The La–Nd samples (pre-compressed 1-μm thick pieces) mixed with ammonia borane $NH_3BH_3$ were heated by a pulsed laser light for several hundred microseconds to 1500–2000 K at pressures of 170–180 GPa (determined via the Raman signal of diamond[15]) in DACs M1–E3 (Supporting Information Table S1). We most fully investigated the properties of lanthanum-neodymium superhydrides prepared from the La–Nd (9 ± 0.5 %) alloy by powder X-ray diffraction (XRD) of synchrotron radiation with a wavelength of 0.413 Å (Figure 1).

We found that a series of the strongest reflections in the XRD pattern (Figure 1a) corresponds to a cubic crystal structure, which is typical for decahydrides such as $LaH_{10}$ at pressures above 150 GPa. The experimental equation of state $V(P)$ of the main phase (Figure 1b) allowed us to determine the metal-to-hydrogen ratio in the obtained compound to be 1:10. Considering good agreement with theoretical calculations and the fact that the initial composition of the La–Nd alloy is close to $La_9Nd$, we further assign to this phase the chemical formula $Fm\bar{3}m$-$(La_{0.91}Nd_{0.09})H_{10}$ with a unit cell volume of ~30.3 Å$^3$ per metal atom at 200 GPa (Supporting Information Table S2), slightly lower than the cell volume of $Fm\bar{3}m$-$LaH_{10}$ (Figure 1b).[6,16] In a recent experiment by N. Salke et al.[17] with $La_{0.82}Nd_{0.18}$ alloy at 170 GPa a similar cubic polyhydride $(La,Nd)H_{10}$ with a unit cell volume of 33.3 Å$^3$/metal atom was obtained.

It is worth pointing out that cubic $NdH_{10}$, which was not obtained during the direct synthesis below 130 GPa,[18] can be stabilized in the $LaH_{10}$ sublattice at higher pressure. We have previously observed the same behavior for $YH_{10}$, stabilized in the lattice of ternary polyhydrides $(La,Y)H_{10}$.[19] The impurity phase in the sample (asterisks in the Figure 1a) has a low hydrogen content and can be explained by the composition $Immm$-$(La,Nd)_3H_8$, which is ~4.4 meV/atom above the convex hull of the La-Nd–H system at 200 GPa.

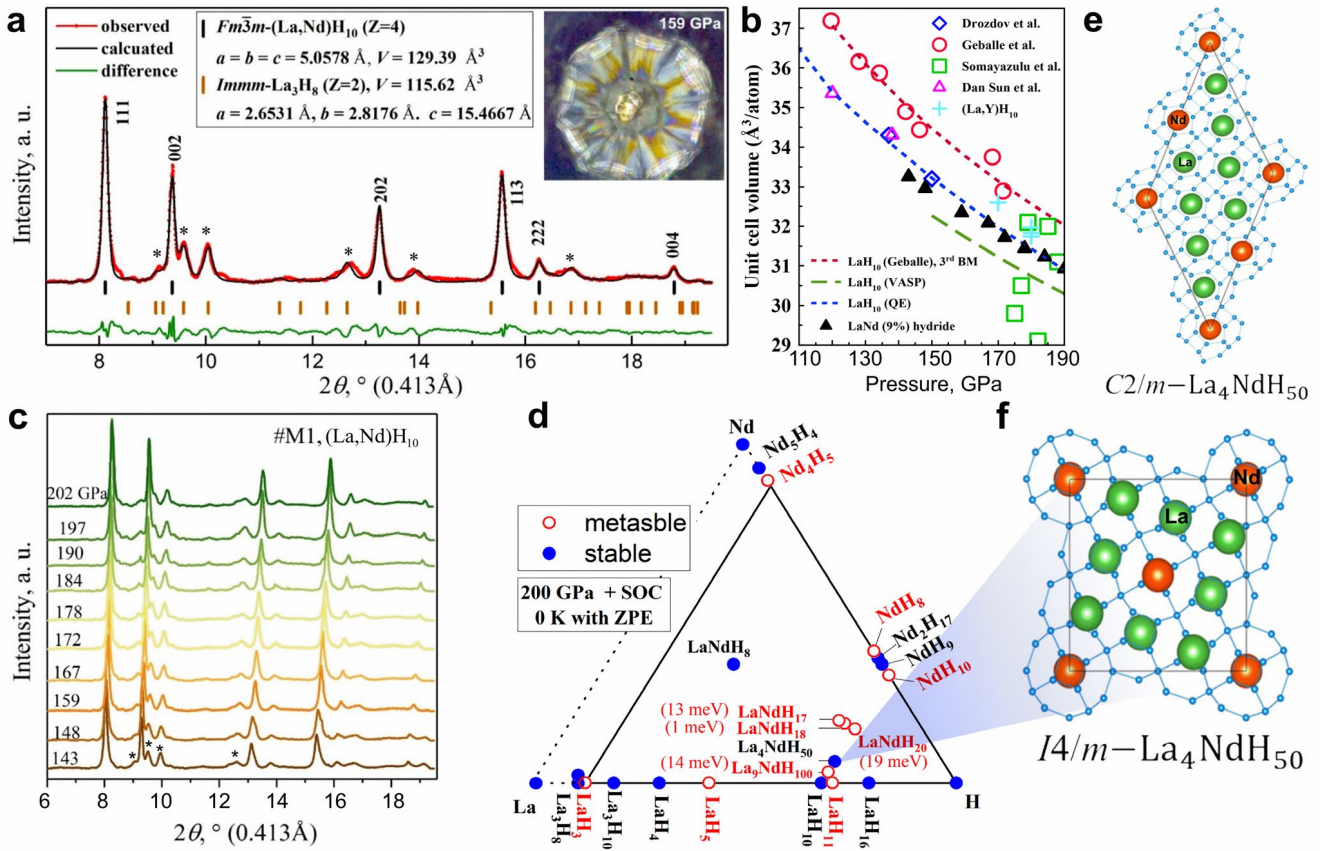

**Fig. 1 XRD experimental and theoretical explanations for (La,Nd)H$_{10}$.** (a) Experimental XRD pattern and the Le Bail refinement of the (La,Nd)H$_{10}$, obtained from La$_{0.91}$Nd$_{0.09}$ alloy, and $Immm$-La$_3$H$_8$ unit cell parameters at 159 GPa. (b) Experimental and theoretical equations of state of LaH$_{10}$[6,16,20,21] and newly synthesized (La,Nd)H$_{10}$. (c) XRD patterns obtained during decompression of DAC M1 from 202 to 143 GPa. (d) Calculated convex hull of the ternary La–Nd–H system at 200 GPa considering the spin–orbit coupling (SOC) and zero-point energy (ZPE) at 0 K. Stable and metastable phases are shown in blue and red, respectively. (e, f) Crystal structures of the model compounds — ternary La–Nd hydrides with pseudocubic (fcc) metal sublattices $I4/m$-La$_4$NdH$_{50}$ and $C2/m$-La$_4$NdH$_{50}$.

In the context of the joint experimental and theoretical approach, we analyzed the stability of various La–Nd–H phases using the USPEX code.[22-24] The most thermodynamically stable ternary polyhydride is $C2/m$-La$_4$NdH$_{50}$, which has a high-symmetry modification $I4/m$-La$_4$NdH$_{50}$ lying 5 meV above the convex hull. Comparing the stability of different (La,Nd)H$_{10}$ phases toward their decomposition to LaH$_{10}$ + NdH$_{10}$, we see that pseudocubic $P1$-La$_9$NdH$_{100}$ and $C2/m$-La$_4$NdH$_{50}$ are stable at 200 GPa (Supporting Information Figure S8a-c). The situation does not change when the zero-point energy (ZPE) is considered (Supporting Information Figures S8d-f), but the results essentially depend on the spin–orbit coupling (SOC) and the type of La pseudopotential used. The calculations that include $Z = 11$ outer shell electrons for La show stabilization of (La,Nd)H$_{10}$ with higher Nd content (Supporting Information Figure S8a-b), which may explain the emergence of impurities of lower La hydrides (e.g., (La,Nd)$_3$H$_8$) in the experiment.

In the harmonic approximation, all model structures of (La,Nd)H$_{10}$ are unstable and have imaginary modes at 200 GPa, just like $Fm\bar{3}m$-LaH$_{10}$ at this pressure[25] (Supporting Information Figure S10). More accurate anharmonic calculations based on molecular dynamics and machine learning interatomic potentials (MLIP[26-29], Supporting Information Figure S12a-d) show that all studied phases — $P\bar{1}$-La$_9$NdH$_{100}$, $P2/m$-La$_9$NdH$_{100}$, $C2/m$-La$_4$NdH$_{50}$, and $I4/m$-La$_4$NdH$_{50}$ — are stable and retain their pseudocubic structures.

Theoretical modeling using true ternary phases of the La–Nd–H system: La$_9$NdH$_{100}$, La$_4$NdH$_{50}$, etc., obtained by simple replacement of La with Nd in supercells of $Fm\bar{3}m$-LaH$_{10}$, is very convenient. Such La–Nd–H phases give correct LaH$_{10}$-like XRD spectra as well as cell volumes and are dynamically and thermodynamically stable. Despite the convenience of these models, the synthesized structure is most likely a solid solution with randomly distributed La and Nd atoms in the metal sublattice of the polyhydride. This conclusion can be made because of the absence of superstructure XRD reflections and a very low barrier to the movement of Nd atoms in the La sublattice (see Supporting Information Figure S9).

**Superconducting properties of La–Nd hydrides**

Superconductivity in $Fm\bar{3}m$-(La$_{0.91}$Nd$_{0.09}$)H$_{10}$ was experimentally investigated at 170–180 GPa in the electrical DACs E0 and E1 (see Table S1 in Supporting Information). This hydride demonstrates relatively wide superconducting transition ($\Delta T_C \approx 30$ K), with the onset superconducting critical temperature $T_C = 148$ K (Figure 2a), about 100 K lower than in pure parent LaH$_{10}$ [9] at the same pressure. The residual resistance at 100 K does not exceed 0.5 mΩ as compared to 0.8 Ω in the normal state at 160 K. In this case, we can consider the resistance $R(T)$ above $T_C$ also comes from the synthesized hydride and use the Bloch–Gruneisen formula[30] which gives the Debye temperature $\theta_D \approx 1270$ K (1156 K according to Ref. [31]) and the corresponding electron–phonon coupling strength $\lambda_{B-G} \approx 1.62$ ($\mu^* = 0.1$).

The superconducting transition was initially studied in steady magnetic fields up to 16 T (Figure 2b). In this range of fields the upper critical magnetic field of (La$_{0.91}$Nd$_{0.09}$)H$_{10}$ shows linear dependence $\mu_0 H_{C2} \sim |T - T_C|$ with the slope $d\mu_0 H_{C2}(T_C)/dT = -0.71$ T/K. The superconducting transitions reveal no significant broadening[32-34] in the range of fields (0–16 T): ($T_C^{90\%} - T_C^{10\%}$) ~ 23–25 K, $\Delta T_C/T_C \sim 0.2$. Different models were used to extrapolate the upper critical magnetic field to 0 K. The Werthamer–Helfand–Hohenberg (WHH) model[35] yields $\mu_0 H_{C2}(0) = 73$ T (Figure 2d). A $(1-t^2)$-model, also called the Ginzburg-Landau (GL) model [36], gives much lower $\mu_0 H_{C2}(0) \sim 54$ T. As we show below, a linear extrapolation yields a result that is much closer to the experiment.

To extend the magnetic phase diagram of (La$_{0.91}$Nd$_{0.09}$)H$_{10}$ and verify its extrapolated value $\mu_0 H_{C2}(0)$, we used pulsed fields up to 68 T (Figure 2c,e) and a special DAC E1 made of Ni–Cr–Al alloy. The dependence of the upper critical magnetic field on temperature is linear up to fields of about 68 T, which is 64–70 % of the extrapolated value $\mu_0H_{C2}(0)$ = 100–110 T (Figure 2d,e). It has recently been shown that some other hydrides (e.g., low-pressure modification of LaH$_{10}$,[37] SnH$_{~4}$,[38,39] YH$_4$[40]) exhibit the linear dependence $\mu_0H_{C2}(T) = a \times (T_C - T)$ in the overall temperature range (see also Discussion). This contradicts the conventional WHH model widely used for hydrides, which predicts saturation of $\mu_0H_{C2}(T)$ in such strong magnetic fields and, hence, underestimates $\mu_0H_{C2}(0)$.

Another important result is the observation of the broadening of superconducting transitions in the first series of experiments with pulsed magnetic fields. The broadening of resistive transitions in magnetic fields is a distinctive feature of most superconductors, but it is not always observed in compressed polyhydrides.[34] The absence of broadening of resistive transitions is also seen in our (La,Nd)H$_{10}$ sample in weak magnetic fields up to 16 T (<16% of $\mu_0H_{C2}(0)$, Figure 2b). However, as seen in Figure 2c, such broadening appears in much stronger magnetic fields $H > 0.5H_{C2}(0)$, which are available only with pulse magnets. More specifically, in the range from 7 to 53 T, the superconducting transitions of (La$_{0.91}$Nd$_{0.09}$)H$_{10}$ demonstrate a notable broadening expected in conventional superconductors, with $T_C^{0.25\Omega} - T_C^{0.025\Omega}$ changing from 14 to 28 K and $\Delta T_C/T_C$ increasing from 12% to 41% (Figure 2c). In the second series (run 2) of high-field measurements, it was not possible to observe the broadening of the superconducting transitions because of their large width for this sample even in the absence of a magnetic field (Supporting Information Figure S19b).

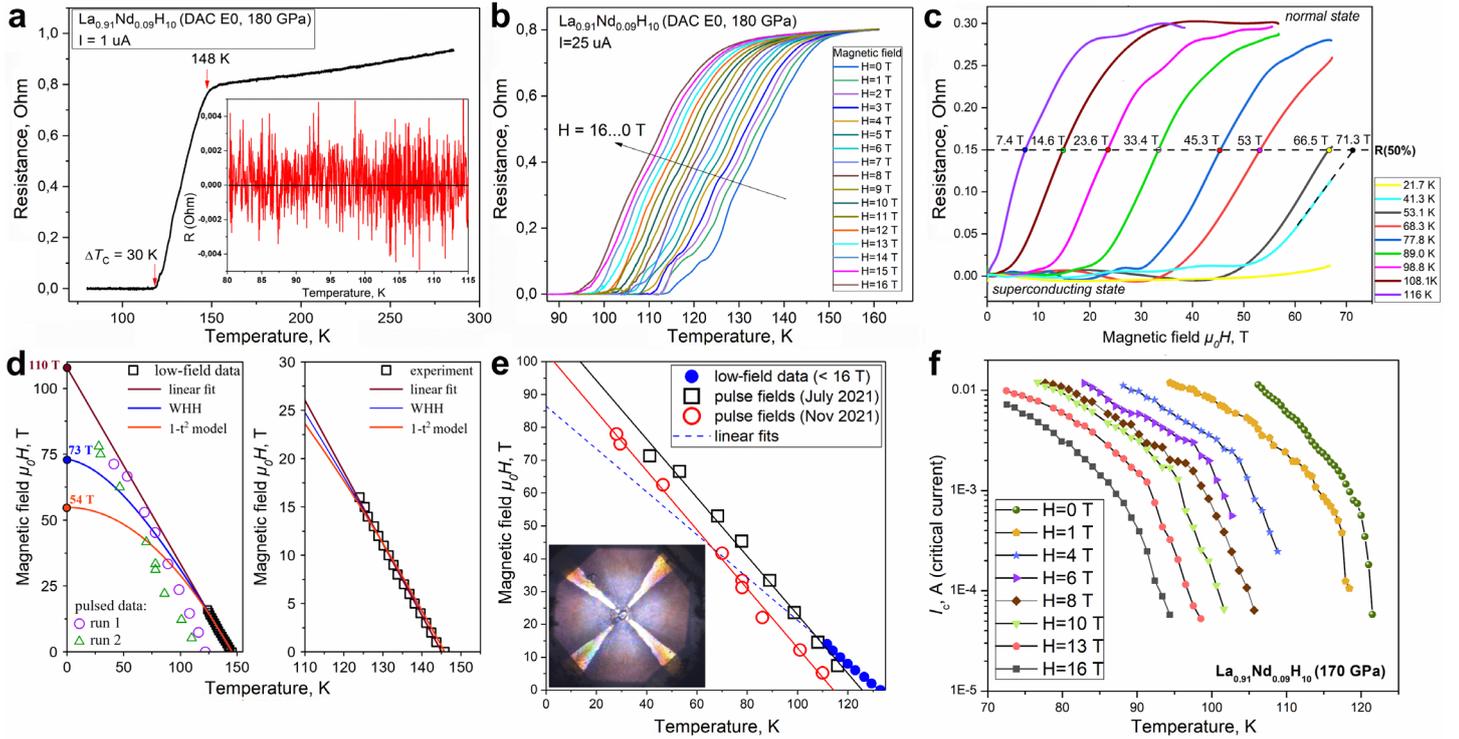

**Fig. 2 High-pressure electrical measurements in La–Nd hydrides in DACs E0 and E1 (180 and 170 GPa).** (a) Temperature dependence of the electrical resistance of the synthesized sample at 180 GPa (DAC E0, the current frequency (*f*) is 4 Hz). Inset: residual resistance in the superconducting state. (b) Superconducting transitions in magnetic fields of 0–16 T. (c) Electrical resistance of the sample in strong pulsed magnetic fields (*f* = 3.33 kHz) at different temperatures (run 1). (d) Extrapolation of the upper critical magnetic field $H_{C2}(0)$ using the Werthamer–Helfand–Hohenberg model (WHH), linear fit, and $(1 - t^2)$ Ginzburg-Landau model. (e) Magnetic phase diagram of (La$_{0.91}$Nd$_{0.09}$)H$_{10}$ synthesized in DAC E1 (170 GPa). Inset: photo of the diamond anvil culet with the sample and electrodes. (f) Dependence of the critical current $I_C(T, B)$ on the applied magnetic field and temperature (DAC E1).

The linear dependence of the upper critical magnetic field on the temperature requires a brief discussion. For BCS superconductors, the generally accepted model of the $\mu_0H_{C2}(T)$ dependence is WHH, which predicts saturation (flattening) of $\mu_0H_{C2}(T)$ at low temperatures[35]. However, for many compressed polyhydrides (YH$_4$, LaH$_{10}$ at low pressures[41]), completely linear $\mu_0H_{C2}(T)$ dependence was observed down to ~1–2 K. This behavior may indicate the presence of two superconducting gaps,[42-46] which is confirmed for a number of polyhydrides ($Fm\bar{3}m$-LaH$_{10}$, $Fm\bar{3}m$-YH$_{10}$, $P6_3/mmc$-YH$_9$, etc.) by solving the anisotropic Migdal–Eliashberg equations.[47,48] However, this explanation seems unsatisfactory, since the two-gap superconductivity is not universal in hydrides (e.g., $Im\bar{3}m$-CaH$_6$[49]), nevertheless, ascending segments of the $\mu_0H_{C2}(T)$ dependence have never been observed, and polyhydrides following the WHH model are unknown at the moment. Almost all superconducting polyhydrides known today exhibit a close to linear $\mu_0H_{C2}(T)$ dependence (Supporting Information Figure S18).

We believe that the linear behavior of $\mu_0H_{C2}(T)$ is associated with the mesoscopic inhomogeneity of the sample, that is the presence of regions with different composition and hydrogen content and, consequently, different values of $T_C$ and $\mu_0H_{C2}$.[50,51] Such inhomogeneous state may be considered as a granular superconductor. As long as "islands" of superconductivity are bound via the Josephson effect, we can detect superconductivity in the sample. The proposed explanation agrees with the current absence of observations of the Meissner effect in hydride superconductors, which is very difficult to detect, whereas the diamagnetic screening (at zero-field cooling) has been found.[16,52]

**Critical current, impurity concentration and magnetoresistance of La–Nd hydrides**

Superconductivity can be suppressed by reaching the critical temperature, the upper critical magnetic field or the critical current density. The critical currents and voltage–current ($U$–$I$) characteristics for the (La$_{0.91}$Nd$_{0.09}$)H$_{10}$ sample were investigated in the range from $10^{-5}$ to $10^{-2}$ A in external magnetic fields at pressure of 170 GPa (DAC E1, Figure 2f). The critical current density was estimated from the size and thickness of the sample, which were 25–30 μm and ~1 μm (the thickness of the loaded piece of the La–Nd alloy), respectively. It was established that the critical current density in (La,Nd)H$_{10}$ exceeds 400 A/mm$^2$ at 105 K. The extrapolation to T=0 K using the conservative Ginzburg–Landau model[36] $J_C = J_C(0) \times (1 - T/T_C)^{3/2}$ gives the critical current $I_C(0) = 0.22$ A and the critical current density $J_C(0)$ that may exceed 8.8 kA/mm$^2$. The single vortex model[53] $J_C = J_C(0) \times (1 - T/T_C)^{5/2}(1 + T/T_C)^{-1/2}$ yields much higher values: the critical current $I_C$ may reach 3 A at 0 K and the critical current density $J_C(0)$ may exceed 120 kA/mm$^2$. However, the heterogeneous structure of the sample can significantly reduce the real values of the current density that can be achieved.

The critical current measurements can be used to estimate the superconducting gap (Figure 3d). Talantsev and Tallon proposed the following model for $s$-wave superconductors:[54]

$$I_C(T)^{2/3} = I_C(0)^{2/3}\left(1 - 2\sqrt{\frac{\pi\Delta(0)}{k_BT}}\exp\left(-\frac{\Delta(0)}{k_BT}\right)\right), \qquad (1)$$

where $I_C(T)$ is the critical current at temperature $T$ in the absence of the magnetic field and $\Delta(0)$ is the superconducting gap at 0 K. Interpolation of the $I_C(T)$ data using both Ginzburg–Landau and Dew-Hughes models for $I_C(0)$ and Eq. (1) yields similar $\Delta(0) = 14.5$–15 meV, which, however, is lower than expected from $2\Delta(0)/k_BT_C = 3.52$.

According to the Gor'kov theory, the $T_C$-dependence for (La,Nd)H$_{10}$ on the Nd concentration ($x \ll 1$) can be expressed as a linear function:[11,55]

$$k_B(T_C^{max} - T_C(x)) = \frac{\pi\hbar}{4\tau}x, \qquad (2)$$

where τ is the collision time resulting from the impurity potential in the Born approximation. In our case, $\tau = 5.4 \times 10^{-15}$ s, $x = 0.09$ (atomic fraction of Nd), which corresponds to a "dirty" metal. Thus, each atomic percent of neodymium suppresses $T_C(LaH_{10})$ by about 10–11 K or, in relative values, $\Delta T_C(1\% \, Nd)/T(LaH_{10}) = 0.044$ (Figure 3e), which is much less than the degree of suppression of superconductivity in pure lanthanum,[56] where $\Delta T_C(1\% \, Nd)/T(La) = 0.144$. This must be considered when synthesizing $LaH_{10}$, where impurities of ~0.1% of other lanthanides can negatively affect the attainable critical temperature.

According to theory[11,55], the critical concentration $x_{cr}$ of a magnetic impurity that completely suppresses superconductivity in $LaH_{10}$ can be derived from the experiment:

$$x_{cr} = \frac{\pi^2}{8\gamma} \frac{T_C^{max}}{(T_C^{max} - T(x))} x, \qquad (3)$$

where $\gamma \approx 1.78$ is the Euler constant; then $x_{cr} = 0.15$. Indeed, $(La,Nd)H_{10}$ synthesized from the $La_{0.8}Nd_{0.2}$ alloy containing 20 at% of neodymium does not demonstrate clear superconducting transitions (DAC E2, Supporting Information Figure S20). In DACs E3 and E4, loaded with $La_{0.92}Nd_{0.08}$ and $La_{0.935}Nd_{0.065}$ alloys, we observed superconducting transitions at 157 K and 179 K, respectively (Supporting Information Figure S21).

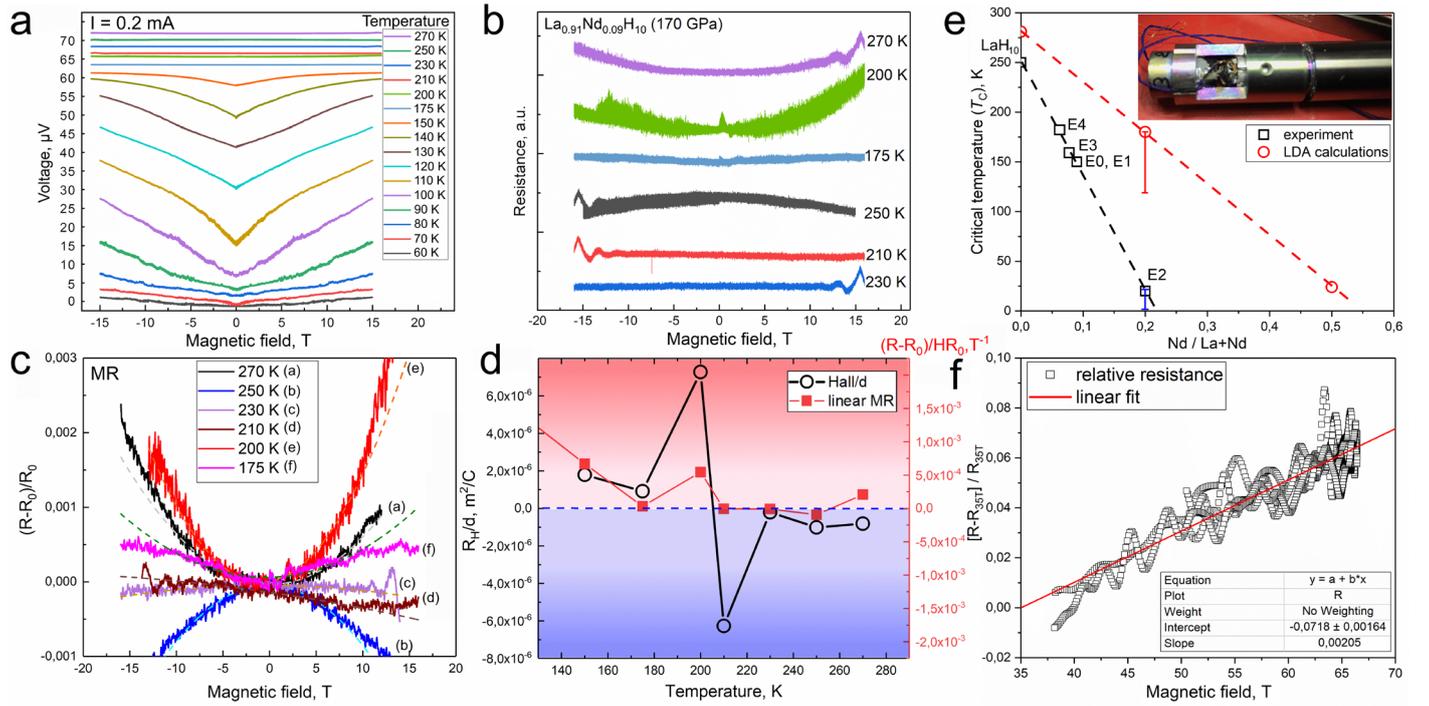

**Fig. 3 Galvanomagnetic effects in $La_{0.91}Nd_{0.09}H_{10}$ at ~170 GPa measured in steady and pulsed magnetic fields.** (a) Dependence of the detected voltage on the applied magnetic field (from –16 to 16 T) at different temperatures. Between 90 and 150 K, the main contribution to the magnetoresistance (MR) is due to the superconducting transition. (b) Dependence of the sample resistance (arbitrary units, the curves are offset-shifted for clarity) on the applied magnetic field for $T > T_C$. There is an obvious change in the sign of MR in 200-250 K region. (c) Detailed comparison of the dependence of the relative resistance on the magnetic field ($\rho \sim H^2$) and its quadratic character at low fields. Dashed lines correspond to parabolic fit. (d) Dependence of the linear term in MR and the Hall coefficient divided by $d$ ($R_H/d$, where $d$ is the poorly known thickness of the sample) at $T > T_C$. (e) Experimental and calculated (using the local density approximation) dependence of $T_C$ on the concentration of Nd in $(La,Nd)H_{10}$ at 170–180 GPa. Inset: sample position in the DAC E1 for measurements of MR in pulsed magnetic fields. (f) Linear term in magnetoresistance in pulsed magnetic fields of 37–67 T at 105±5 K.

Low-field measurements in (La,Nd)H$_{10}$ doped by 9 at% of Nd (Figure 3a) indicate the presence of two terms in the field dependencies of the magnetoresistance (MR, Figure 3c): quadratic ρ = (R-R$_0$)/R$_0$ = μ$^2$H$^2$ (dominating in low fields), where μ – is the electron mobility (Table 1), and linear ρ = αH (dominating in fields above 7–9 T). In the experiment with a pulsed magnetic field, a trend of a linear resistance increase with the field is seen despite the noise (Figure 3f). The linear behavior of the magnetoresistance is typical for polycrystalline metals with an open Fermi surface (e.g., spherical with "necks"), for example, Li, Cu, Ag, Au, SrZnSb$_2$[57], MnBi[58] and others.[59] This was explained by Lifshitz and Peschanskii in 1959 by considering the shape of the Fermi surface and the anisotropy of the MR. It leads to a linear dependence of the MR on the field when averaged over all directions (Δθ ~ H$_0$/H, here H$_0$ - is the field for which the cyclotron time equals the electron mean free time) and open sections of the Fermi surface (where ρ ~ H$^2$) for polycrystals.[60] For materials with a closed FS, linear MR is observed in disordered polycristalline narrow-band semiconductors and semimetals. It is also intrinsic to topological insulators and semimetals with linear dispersing Dirac cones.

The configuration of the electrical contacts didn't allow to fully separate the Hall and diagonal voltage drops, and, therefore, the measured voltage drop contained an admixture of the Hall voltage to the diagonal one. The Hall component was then extracted from the asymmetry of the quadratic dependence of the MR in low fields. We found an anomaly in the behavior of the sample (DAC E1) around 200-250 K. The Hall coefficient (Table 1) changes sign in this temperature interval (Figure 3b-d). At the same time, magnetoresistance becomes very small and even negative at 250 K (Figure 3b). Both anomalies are correlated and clearly indicate the band structure and, possibly, crystal structure changes in this temperature range.

The properties of the (La,Nd)H$_{10}$ sample in DAC E1 at low temperature are also unusual. Studying this sample, we found anomaly at 9 K (Supporting Information Figure S22). The electrical resistance begins to decrease sharply at 150 K and reaches a plateau at about 50 K, which corresponds to the superconducting state. Upon cooling below 9 K, the resistance begins to rise again, forming a low-temperature anomaly that does not depend on either the frequency or the excitation current, but can be suppressed by an external magnetic field of ~9 T. The most likely reason for this behavior is a change in the magnetic ordering in the Nd sublattice of (La,Nd)H$_{10}$. Namely, low-temperature jump in electrical resistance may be due to the antiferromagnetic ordering in (La,Nd)H$_{10}$ below the Néel temperature $T_N$ = 12 K, which was predicted using the DFT calculations for simplified $P\bar{1}$-La$_9$NdH$_{100}$ model (see Supporting Information Figures S22-S24).

**Table 1.** Parameters of the normal state of the La–Nd polyhydride (9 at% of Nd) obtained in the study of magnetoresistance in steady and pulsed magnetic fields.

| Temperature, K | Electron mobility, 10$^{-3}$ m$^2$/s×V | Electron relaxation time, 10$^{-14}$ s | $R_H$/d, 10$^{-6}$ m$^2$/C | Hall coefficient $R_H$ at d = 1 μm, 10$^{-12}$ m$^3$/C | Linear MR, (dρ/dH), 10$^{-5}$ T$^{-1}$ |
|---|---|---|---|---|---|
| 105 | – | – | – | – | 200 |
| 150 | ~4.7 | ~2.70 | ~1.79 | ~1.79 | ~66.9 |
| 175 | 1.9 | 1.07 | 0.92 | 0.92 | 3.10 |
| 200 | 3.5 | 1.99 | 7.27 | 0.73 | 54.4 |
| 210 | 0.55 | 0.32 | –6.27 | –6.25 | –0.63 |
| 230 | 0.8 | 0.45 | –0.21 | –0.21 | –0.79 |
| 250 | 2.8 | 1.58 | –1.01 | –1.015 | –9.64 |
| 270 | 2.6 | 1.46 | –0.82 | –0.815 | 20.8 |

* C is Coulomb

The mobility of electrons and their relaxation time (Table 1) in (La,Nd)H$_{10}$, calculated from the quadratic part of the ρ(H) ∝μ$^2$H$^2$ dependence, correspond to the parameters of typical metals.[61] Measurements in low magnetic fields allow to estimate the Hall coefficient $R_H$ at $\sim 10^{-12}$ m$^3$/C (sample thickness $\sim$1–2 μm). Within an order of magnitude, this value of $R_H$ corresponds to those of ordinary metals, which suggests that the concentration of charge carriers in La–Nd hydrides is $\sim 10^{30}$ electron/m$^3$.

## Discussion

The main goal of studying ternary hydrides is to find a way to control the superconducting properties of known polyhydrides (such as $H_3S$, $LaH_{10}$, $YH_6$, $YH_9$, and $CeH_9$) by doping. Nowadays, main hopes for increasing $T_C$ are pinned on ternary and higher order hydrides. However, their studies require complex and accurate calculations, presumably for this reason, many discrepancies between theoretical and experimental[33,62-65] results and overestimations of the superconducting properties have been published in the past few years.[25,62,66-68]

Studies of the effect of nonmagnetic and magnetic impurities on superconductors enabled the distinction between isotropic and anisotropic superconductivity.[11,69] Nonmagnetic (scalar) impurities do not affect the isotropic singlet *s*-wave order parameter (for example, small carbon impurities in lanthanum do not affect the superconducting transition in (La, C)H$_{10}$ that is observed at about 245 K[70]). The absence of the influence of C and Al impurities[70] also casts doubt on attempts to explain the supposedly huge increase in $T_C$ by doping in experiments with C-doped $H_3S$[71] and $LaH_{10}$ doped with Ga, B/N.[65,68,72] One might think that the point here is the formation of true ternary and quaternary polyhydrides. But a detailed theoretical analysis shows[68,73] that this can hardly lead to high-$T_C$ superconductivity in the framework of the BCS theory. Experimental results[65] also do not allow us to draw unambiguous conclusions. Moreover, as we have shown previously, no significant increase in $T_C$ is observed in the La–Y system, which can be considered as the absence of the influence of nonmagnetic yttrium impurities on the superconducting properties of $LaH_{10}$.[19]

This corresponds to so-called Anderson's theorem,[12,74] which states that BCS superconductors are practically insensitive to small amount of nonmagnetic impurities. If the leading mechanism in compressed polyhydrides is the electron–phonon interaction, as most experimental and theoretical studies suggest,[75,76] then it is impossible to expect a significant effect of small additives, such as C or $CH_4$ in $H_3S$,[77,78] on superconductivity. The opposite situation would be a strong argument in favor of an unconventional pairing mechanism in these compounds. As this study shows, La–Nd hydrides exhibit the properties of typical metals, have a very low Hall coefficient, a rather high electron mobility, and a linear term in the magnetoresistance, indicating an open topology of the Fermi surface, which is confirmed by the DFT calculations.[66,79] As we have shown, the observed superconducting and normal state properties are explained within conventional concepts, leaving little chances for unconventional superconductivity in polyhydrides.

In sharp contrast, magnetic scattering is very efficient in destroying *s*-wave superconductivity, as shown in this study. The main observation of this work is that Nd effectively suppresses superconductivity in $LaH_{10}$, whereas its $Fm\bar{3}m$ crystal structure remains almost unchanged because of the great similarity of the physical and chemical properties of La and Nd atoms. With available pulsed magnetic fields, we suppressed superconductivity in (La,Nd)H$_{10}$ much effectively on the $h = B/B_{C2}(0)$ scale than it is currently possible for pure $LaH_{10}$. Given such a strong suppression of superconductivity in (La,Nd)H$_{10}$, where the critical concentration of Nd is $\sim$15 at%, the absence of the superconducting properties in $NdH_9$,[18] $PrH_9$,[80] and in $EuH_9$[81] becomes not surprising anymore. It is hard to believe that high-$T_C$ superconductivity above 200 K can be realized in binary polyhydrides of most other lanthanides (Tm, Yb, Lu), as was recently predicted.[82]

# Conclusions

Superhydrides are a new class of hydrogen-rich materials whose research is a novel direction in materials science. What is the mobility of hydrogen in hydrides at high pressures and temperatures? Are there a photovoltaic effect and photoconductivity? How do photons affect the superconducting transition in hydrides? What are the Hall coefficient and the magnetoresistance? Are they subjected to the topological Lifshitz transition? All these questions have yet to be answered.

In this research, we made a step forward in this direction and successfully synthesized novel ternary polyhydrides (La,Nd)$H_{10}$ containing 8–20 at% of Nd atoms, randomly distributed in LaH$_{10}$-like metal sublattice. The electric transport measurements demonstrated that the addition of magnetic impurities (Nd) leads to a significant suppression of superconductivity in LaH$_{10}$: each atomic % of Nd causes decrease in $T_C$ by 10–11 K. Superconductivity disappears at a critical concentration of Nd of about 15–20 at%.

Using strong pulsed magnetic fields of up to 68 T, we constructed the magnetic phase diagram of synthesized (La,Nd)$H_{10}$, studied the magnetoresistance and the Hall effect. Surprisingly, the dependence of the upper critical magnetic field on temperature is found to be completely linear, possibly due to the mesoscopic inhomogeneity of the hydride sample. The extrapolated upper critical magnetic field $\mu_0 H_{C2}(0)$ for La$_{0.91}$Nd$_{0.09}$H$_{10}$ is 105–110 T at 170 GPa. The current-voltage measurements showed that the critical current density $J_C(0)$ in (La,Nd)$H_{10}$ may exceed 8.8 kA/mm$^2$.

In this study, we showed that La–Nd hydrides exhibit the properties of normal metals: they have a rather small Hall coefficient ($R_H \sim 10^{-12}$ m$^3$/C), high concentration of charge carriers of $\sim 10^{30}$ electron/m$^3$, and a linear part of the magnetoresistance, indicating an open topology of the Fermi surface and the presence of Dirac cones in the vicinity of the Fermi energy[79]. The validity of Anderson's theorem for the studied hydrides and the typical-metal galvanomagnetic properties provide strong evidence for the conventional electron-phonon mechanism of superconductivity in hydrogen-rich materials under pressure.

# Author contributions

# Acknowledgments

In situ X-ray diffraction experiments at high pressure were performed on SPring-8, station BL10XU, Sayo, Japan (proposal No. 2020A0576). Low-pressure studies were carried out on a synchrotron source of the Kurchatov institute (KISI-Kurchatov), station RKFM. This work was supported by JSPS KAKENHI Grant Number 20H05644.The high-pressure experiments were supported by the Ministry of Science and Higher Education of the Russian Federation within the state assignment of the FSRC Crystallography and Photonics of the RAS and by the Russian Science Foundation (Project No. 19-12-00414). A.R.O. thanks the Russian Science Foundation (grant 19-72-30043). D.V.S. thanks the Russian Foundation for Basic Research (project 20-32-90099). I.A.K. thanks the Russian Science Foundation (grant No. 21-73-10261) for the financial support of the anharmonic phonon density of states calculations and molecular dynamics simulations. SEM, XRF, and XRD studies of the initial alloys were performed using the equipment of the Shared Research Center FSRC Crystallography and Photonics of the RAS. I.A.T. and A.G.I. acknowledge the use of the facilities of the Center for Collective Use "Accelerator Center for Neutron Research of the Structure of Substance and Nuclear Medicine" of the INR RAS for high-pressure cell preparation. The research used resources of the LPI Shared Facility Center. V.M.P, A.V.S. and O.A.S. acknowledge the support of the state assignment of the Ministry of Science and Higher Education of the Russian Federation (Project No. 0023-2019-0005) and the Russian Science Foundation, grant 22-22-00570. K.S.P. thanks the Russian Foundation for Basic Research (project 19-02-00888). I.A.K. thanks the Russian Science Foundation (grant No. 19-73-00237) for the financial support of the development of T-USPEX method and anharmonic phonon density of states calculation algorithm. S.W.T was supported by NSF Cooperative Agreement No. DMR-1157490/1644779 and by the State of Florida. A.D.G. was supported by T.H. & T.F. funding. We acknowledge the support of the HLD at HZDR, member of the European Magnetic Field Laboratory (EMFL). We also thank Igor Grishin (Skoltech) for proofreading the manuscript, and Christian Tantardini for calculations using the virtual crystal approximation.

## Data availability

The raw and processed data required to reproduce these findings are available to download from GitHub: https://github.com/mark6871/SPring-8-February-2021-, https://github.com/mark6871/-La-Nd-H10_Transport_Measurements, and as Supporting Information for the manuscript.

# SUPPORTING INFORMATION

# Effect of paramagnetic impurities on superconductivity in polyhydrides:

# $s$-wave order parameter in Nd-doped LaH$_{10}$

Dmitrii V. Semenok,[1,*] Ivan A. Troyan,[2] Andrey V. Sadakov,[5] Di Zhou,[1,*] Michele Galasso,[1] Alexander G. Kvashnin,[1] Ivan A. Kruglov,[8,9] Alexey A. Bykov,[3] Konstantin Y. Terent'ev,[4] Alexander V. Cherepahin,[4] Oleg A. Sobolevskiy,[5] Kirill S. Pervakov,[5] Alexey Yu. Seregin,[2,13] Toni Helm,[12] Tobias Förster,[12] Audrey D. Grockowiak,[10, 11] Stanley W. Tozer,[11] Yuki Nakamoto,[7] Katsuya Shimizu,[7] Vladimir M. Pudalov,[5,6] Igor S. Lyubutin,[2] and Artem R. Oganov[1,*]

[1] Skolkovo Institute of Science and Technology, 121205, Bolshoy Boulevard 30, bld. 1, Moscow, Russia

[2] Shubnikov Institute of Crystallography, Federal Scientific Research Center Crystallography and Photonics, Russian Academy of Sciences, 59 Leninsky Prospekt, Moscow 119333, Russia

[3] NRC "Kurchatov Institute" PNPI, Gatchina, Russia

[4] Kirensky Institute of Physics, Krasnoyarsk, Russia

[5] P. N. Lebedev Physical Institute, Russian Academy of Sciences, Moscow 119991, Russia

[6] National Research University Higher School of Economics, Moscow 101000, Russia

[7] KYOKUGEN, Graduate School of Engineering Science, Osaka University, Machikaneyamacho 1-3, Toyonaka, Osaka, 560-8531, Japan

[8] Dukhov Research Institute of Automatics (VNIIA), Moscow 127055, Russia

[9] Moscow Institute of Physics and Technology, 9 Institutsky Lane, Dolgoprudny 141700, Russia

[10] Brazilian Synchrotron Light Laboratory (LNLS/Sirius), Brazilian Center for Research in Energy and Materials (CNPEM), Campinas, Brazil

[11] National High Magnetic Field Laboratory, Florida State University, Tallahassee, Florida, 32310, USA

[12] Hochfeld-Magnetlabor Dresden (HLD-EMFL), Helmholtz-Zentrum Dresden-Rossendorf (HZDR), 01328 Dresden, Germany

[13] National Research Center "Kurchatov Institute", Moscow 123182, Russia

*Corresponding authors: Artem R. Oganov (a.oganov@skoltech.ru), Di Zhou (D.Zhou@skoltech.ru), Dmitrii Semenok (dmitrii.semenok@skoltech.ru)

# Contents

## Methods

### Experiment

To synthesize the $La_{0.91}Nd_{0.09}$ alloy, La ingots and Nd powder were weighed in the stoichiometric proportion 10:1 in an inert glove box and placed in a $ZrO_2$ crucible filled with toluene to prevent contact with atmospheric oxygen during heating. The crucible was heated to 1000 °C and kept at this temperature for four hours in an induction furnace in a controlled Ar atmosphere. After cooling, the composition of the melted ingot was analyzed using the XRD, X-ray fluorescence (XRF), and EDS methods. The energy-dispersive analysis showed that the obtained alloy contains about 8.5–9.5 at% of Nd. The measurements of the X-ray energy-dispersive spectra (EDS) were performed on FEI Quanta 200 3D scanning electron microscope (SEM) with EDAX Genesys setup.

To prepare the $La_{0.92}Nd_{0.08}$ and $La_{0.8}Nd_{0.2}$ alloys, pure La and Nd (99.9%, CHEMCRAFT Ltd.) were crushed, washed in dilute HCl and acetone to remove impurities, and dried in a glove box. The components were weighed and mixed in a specified ratio. Heating was carried out resistively. The melt was kept in a tantalum crucible at a temperature of 1900 K in an inert atmosphere (argon) for 10 min and quenched at an initial rate of 200 K/min.

To synthesize the $La_{0.74}Nd_{0.26}$ and $La_{0.53}Nd_{0.47}$ alloys, commercially available metallic La and Nd (99.9% purity) produced by Uralredmet Jsc and iTasco Ltd were placed into a glove box with an inert atmosphere. The metals were weighed in the respective stoichiometric ratios and placed in a $ZrO_2$-$Y_2O_3$ crucible, which was filled with toluene and placed into an induction furnace in an inert Ar atmosphere. For $La_{0.74}Nd_{0.26}$, the temperature in the furnace was increased to 950 °C over eight hours and kept for 30 minutes; for $La_{0.53}Nd_{0.47}$, the temperature was raised to 1050°C and kept for 20 minutes. After cooling, samples were taken using a carbide cutter under toluene. The EDS analysis showed the presence of less than 0.1% of yttrium and zirconium from the crucible in the samples.

To load the high-pressure diamond anvil cells (DACs), we took material from homogeneous regions of La-Nd alloys with the desired La:Nd ratio, determined using the EDS and XRF methods. We used the diamond anvils with a 50 μm culet beveled to 300 μm at 8.5°, equipped with four ~200 nm thick Ta electrodes with ~80 nm gold plating that were sputtered onto the piston diamond. Composite gaskets consisting of a rhenium ring and a $CaF_2$/epoxy mixture were used to isolate the electrical leads. Lanthanum–neodymium pieces with a thickness of ~1–2 μm were sandwiched between the electrodes and ammonia borane $NH_3BH_3$ (AB) in the gasket hole with a diameter of 20 μm and a thickness of 10–12 μm. The laser heating of the samples above 1500 K at pressures of 170–180 GPa by several 100 μs pulses led to the formation of ternary lanthanum–neodymium hydrides whose structure was analyzed using the X-ray diffraction (XRD).

The XRD patterns of the $La_{0.9}Nd_{0.1}H_{10}$ sample were recorded at the BL10XU beamline (SPring-8, Japan) using monochromatic synchrotron radiation and an imaging plate detector at room temperature.[1,2] The X-ray beam with a wavelength of 0.413 Å was focused in a 3 × 8 μm spot with a polymer refractive lens SU-8 produced by ANKA. The XRD data were analyzed and integrated using Dioptas software package (version 0.5).[3] The full profile analysis of the diffraction patterns and the calculation of the unit cell parameters were performed using JANA2006 software[4] with the Le Bail method.[5] The pressure in the DACs was determined via the Raman signal of diamond at room temperature.[6]

Magnetoresistance measurements under high magnetic fields were conducted in the 24 mm bore, 72 T resistive pulse magnet (rise time of 15 ms) at the Helmholtz-Zentrum Dresden-Rossendorf (HZDR). Two series of measurements were performed: in July 2021 (run 1) and in November 2021 (run 2). In the first series of experiments, a He bath cryostat and a NiCr coil (heater) wound on a diamond cell were used. For temperatures above 77 K, only helium gas was used in the cryostat, lower temperatures were reached by immersing the capsule in liquid helium. Silver paint Silberleitlack LS200N (Keramikbedarf) was used to lengthen the electrical contacts deposited on the diamond anvils. Strands of the Litz wire glued to the silver paint were moved closer together to minimize open loop pickup. The DAC was encased with strips of 125 μm × 1 cm wide Kapton tape. All twisted pairs were fixed using the GE7031 varnish (Lake Shore, 50:50 toluene:methanol thinner). Wiring was brought out through four holes in a lid of the VTI (Variable Temperature Insert) can which were opened to ∅2 mm. The 50 Ω heater was wrapped over the Kapton tape on the DAC nut (HPN insulated California Fine Wire Stableohm 650∅125 μm) and connected to upper pads just below the T-sensor. 100 μm Cu wire was connected to the heater wires in the gap between the nut and the cell body. The T-sensor (Lake Shore Cernox X95809) was secured to the flat of the DAC opening at the height of the sample wiring connected to pads above the heater.

In the second series of experiments, a flow He cryostat (VTI) was used, which made it possible to better control the DACs temperature. Cernox thermometers were attached to the DAC gasket for accurate measurements of the sample temperature. There was no observable heating from the ramping of the magnetic field at rates up to 100 T/sec above 20 K. A high-frequency (3.33 and 33.3 kHz) lock-in amplifier technique was employed to measure the sample resistance in a 72 T pulsed magnet. The magnet can be used on very special occasions to 72 T, but is usually used to 65 T to extend its lifetime. For the measurements in high magnetic fields, we used a four-wire AC method with the excitation current of 0.5–1 mA;[7] the voltage drop across the sample was amplified by an instrumentation amplifier and detected by a lock-in amplifier. No sample heating was observed during ~150 ms long magnet pulse at temperatures above 20 K from comparisons of up sweep and down sweep resistance traces at different field sweep rates. Numerous heating and cooling cycles, during which the pressure in DACs changes, led to mechanical displacement of a sample, as well as to changes in the hydrogen concentration in different parts of it. This process is similar to the deformation embrittlement and aging of metals. Indeed, some degradation of the sample and broadening of the superconducting transition in it with the emergence of additional steps were observed. For this reason, the $(La,Nd)H_{10}$ samples in the first and second series of experiments slightly differ in their properties.

**Theory**

The computational predictions of thermodynamic stability of the La–Nd–H phases at 200 GPa were carried out using the variable-composition evolutionary algorithm USPEX.[8-10] The first generation consisting of 80 structures was produced using the random symmetric[10] and random topology[11] generators, whereas all subsequent generations contained 20% of random structures and 80% of those created using heredity, softmutation, and transmutation operators. The evolutionary searches were combined with structure relaxations using the density functional theory (DFT)[12,13] within the Perdew–Burke–Ernzerhof (PBE) generalized gradient approximation (GGA) functional[14] and the projector augmented wave method[15,16] as implemented in the VASP code.[17-19] The kinetic energy cutoff for plane waves was 600 eV. The Brillouin zone was sampled using Γ-centered $k$-points meshes with a resolution of $2\pi \times 0.05$ Å$^{-1}$. The same parameters were used to calculate the equations of state of the discovered phases. We also calculated the phonon densities of states of the studied materials using the finite displacements method (VASP and PHONOPY[20,21]). This methodology is similar to the one used in our previous works.[22,23]

The calculations of the critical temperature of superconductivity $T_C$ were carried out using Quantum ESPRESSO (QE) package.[24,25] The phonon frequencies and electron–phonon coupling (EPC) coefficients were computed using the density functional perturbation theory,[26] employing the plane-wave pseudopotential method and the PBE exchange–correlation functional. The critical temperature ($T_C$) was calculated by using the Allen–Dynes formula.[27]

The dynamic stability and phonon density of states of $La_9NdH_{100}$ and $La_4NdH_{50}$ were studied using classical molecular dynamics and interatomic potentials (MTP) based on machine learning.[28] It was demonstrated that the MTP can be used to calculate the phonon properties of materials.[29] Moreover, within this approach we can explicitly take into account the anharmonicity of hydrogen vibrations.

To train the potential, we first simulated $La_9NdH_{100}$ and $La_4NdH_{50}$ in quantum molecular dynamics in an NPT ensemble at 2000 K, with a duration of 5 picoseconds using the VASP code.[17-19] We used the PAW PBE pseudopotentials for the H, La and Nd atoms, $2\pi \times 0.06$ Å$^{-1}$ k-mesh with a cutoff energy of 400 eV. For training of the MTP, a set of $(La,Nd)H_{10}$ structures was chosen using active learning.[30] We checked the dynamical stability of $La_9NdH_{100}$ and $La_4NdH_{50}$ with the obtained MTPs via several runs of molecular dynamics calculations at 300 K. First, the NPT dynamics simulations were performed in a supercell with about 1000 atoms for 40 picoseconds. During the last 20 picoseconds, the cell parameters were averaged. At the second step, the coordinates of the atoms were averaged within the NVT dynamics with a duration of 20 picoseconds and the final structure was symmetrized.

Then, the phonon density of states was calculated within the MTP using the velocity autocorrelator (VACF) separately for each type of atoms:[31]

$$g(\vartheta) = 4 \int_0^\infty \cos(2\pi\vartheta t) \frac{\overline{\langle \vartheta(0)\vartheta(t) \rangle}}{\overline{\langle \vartheta(0)^2 \rangle}} dt \qquad (S1)$$

where $\vartheta$ is the frequency. The velocity autocorrelator was calculated using molecular dynamics, then the phonon DOS was obtained.

# Structural characterization of the initial La–Nd alloys

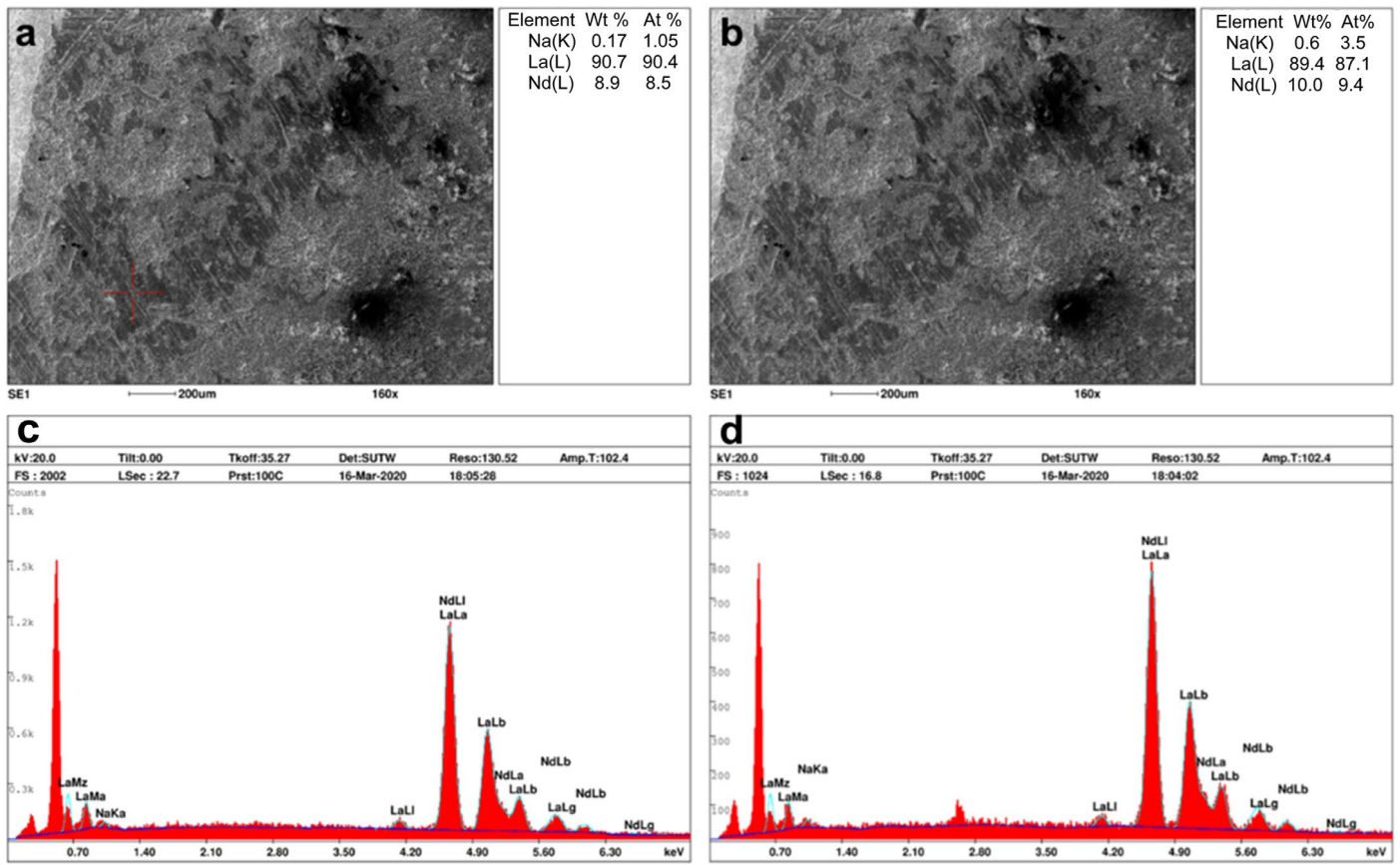

**Figure S1.** (a, b) SEM images of a La–Nd alloy at 0 GPa. Red cross marks the region of the EDS analysis. (c, d) EDS spectra from two different points of the sample. Results show about 9 at% of Nd in the obtained alloy.

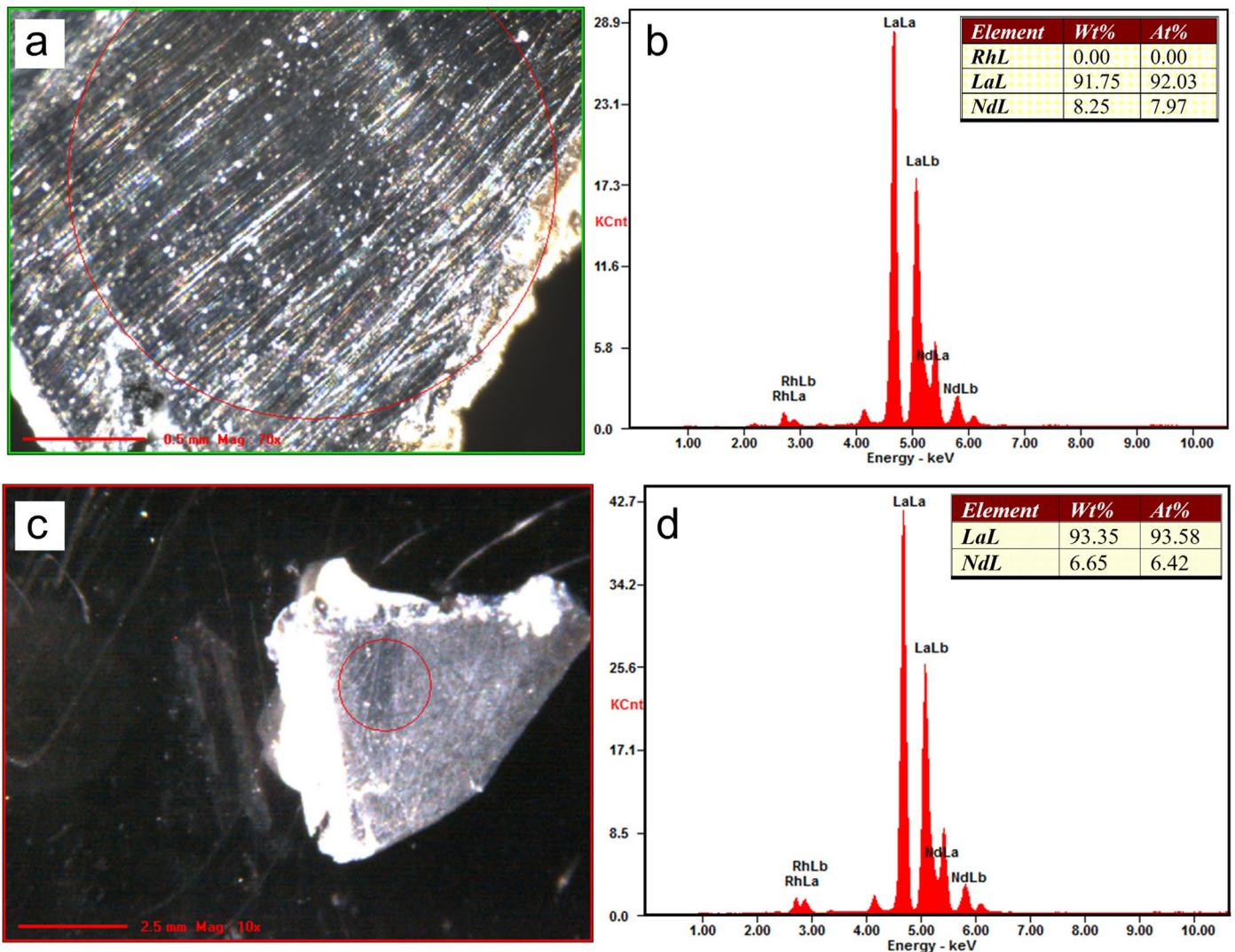

**Figure S2.** SEM image and the EDS analysis of La–Nd alloys at 0 GPa. Results show about 8 at% of Nd (a, b) and 6.5 at% of Nd (c, d) in the obtained alloys. These alloys were used to load DACs E3 and E4.

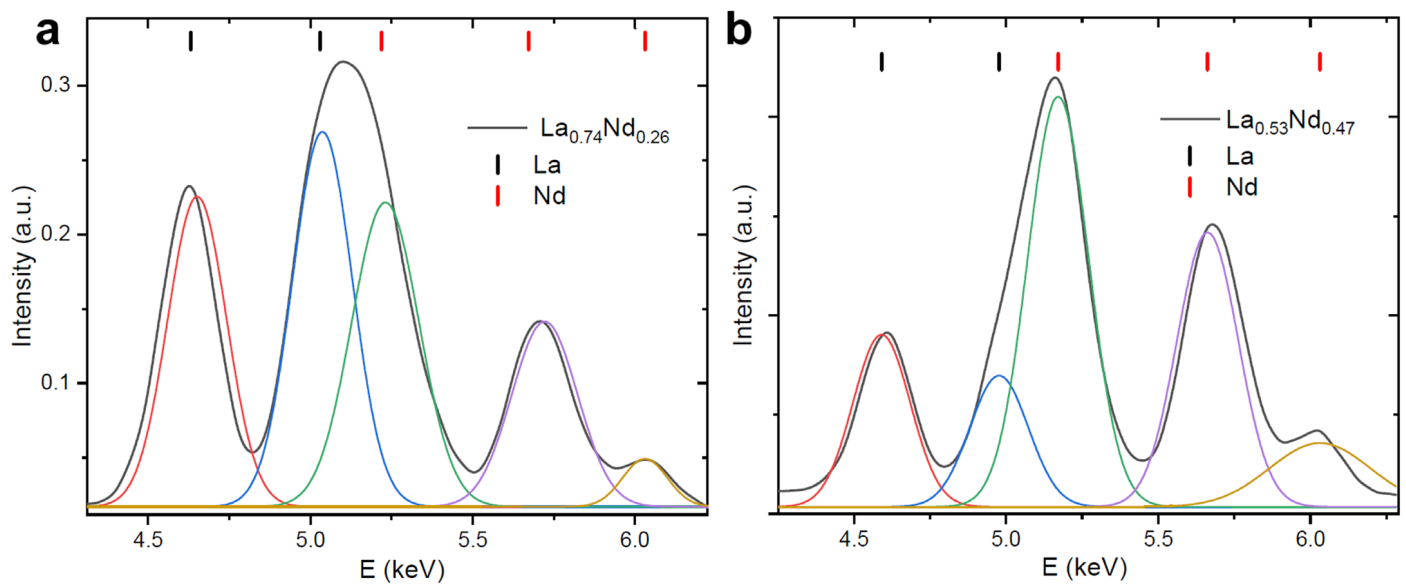

**Figure S3.** Energy-dispersive X-ray spectroscopy (EDS) of the (a) La$_{0.74}$Nd$_{0.26}$ and (b) La$_{0.53}$Nd$_{0.47}$ alloys.

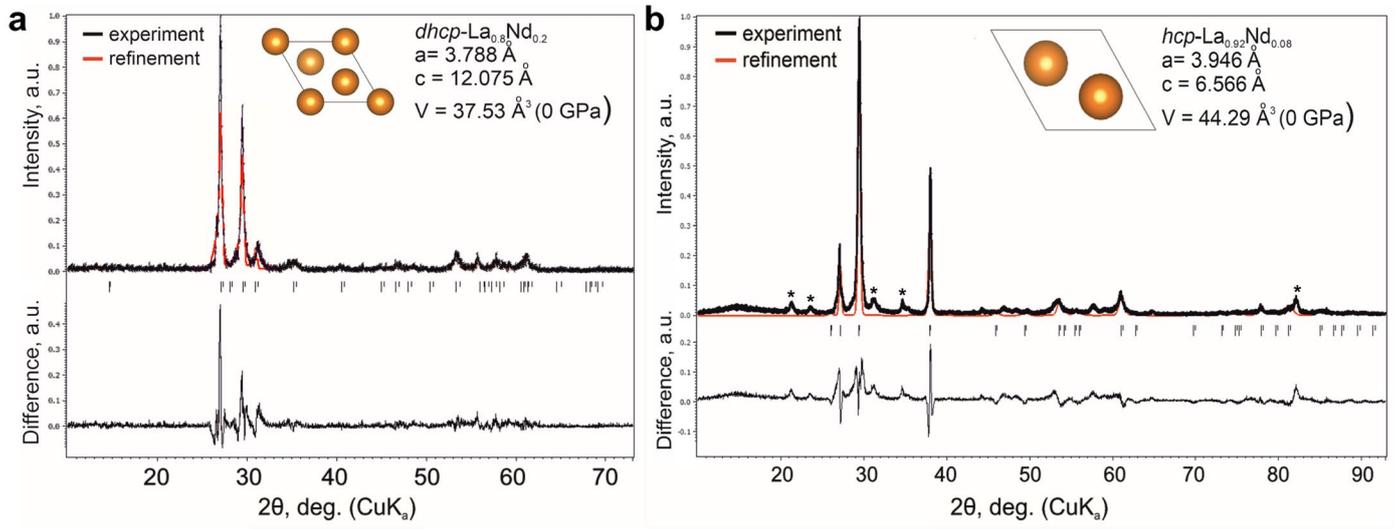

**Figure S4.** X-ray diffraction patterns of the La–Nd alloys at 0 GPa (Cu-K$_a$): (a) *dhcp*-La$_{0.8}$Nd$_{0.2}$, (b) *hcp*-La$_{0.95}$Nd$_{0.05}$. Asterisks indicate uninterpreted reflections.

# Diamond anvil cell characterization

**Table S1.** Experimental parameters of the DACs used in the X-ray diffraction studies. The data in the second column shows the pressure of the laser-assisted synthesis.

| DAC | Pressure, GPa | Composition |
| --- | --- | --- |
| M1 | 202–143 | $La_{0.91}Nd_{0.09}/BH_3NH_3$ |
| E0 | 180 | $La_{0.91}Nd_{0.09}/BH_3NH_3$ |
| E1 | 170–180 | $La_{0.91}Nd_{0.09}/BH_3NH_3$ |
| E2 | 170 | $La_{0.8}Nd_{0.2}/BH_3NH_3$ |
| E3 | 175 | $La_{0.92}Nd_{0.08}/BH_3NH_3$ |
| E4 | 172 | $La_{0.935}Nd_{0.065}/BH_3NH_3$ |

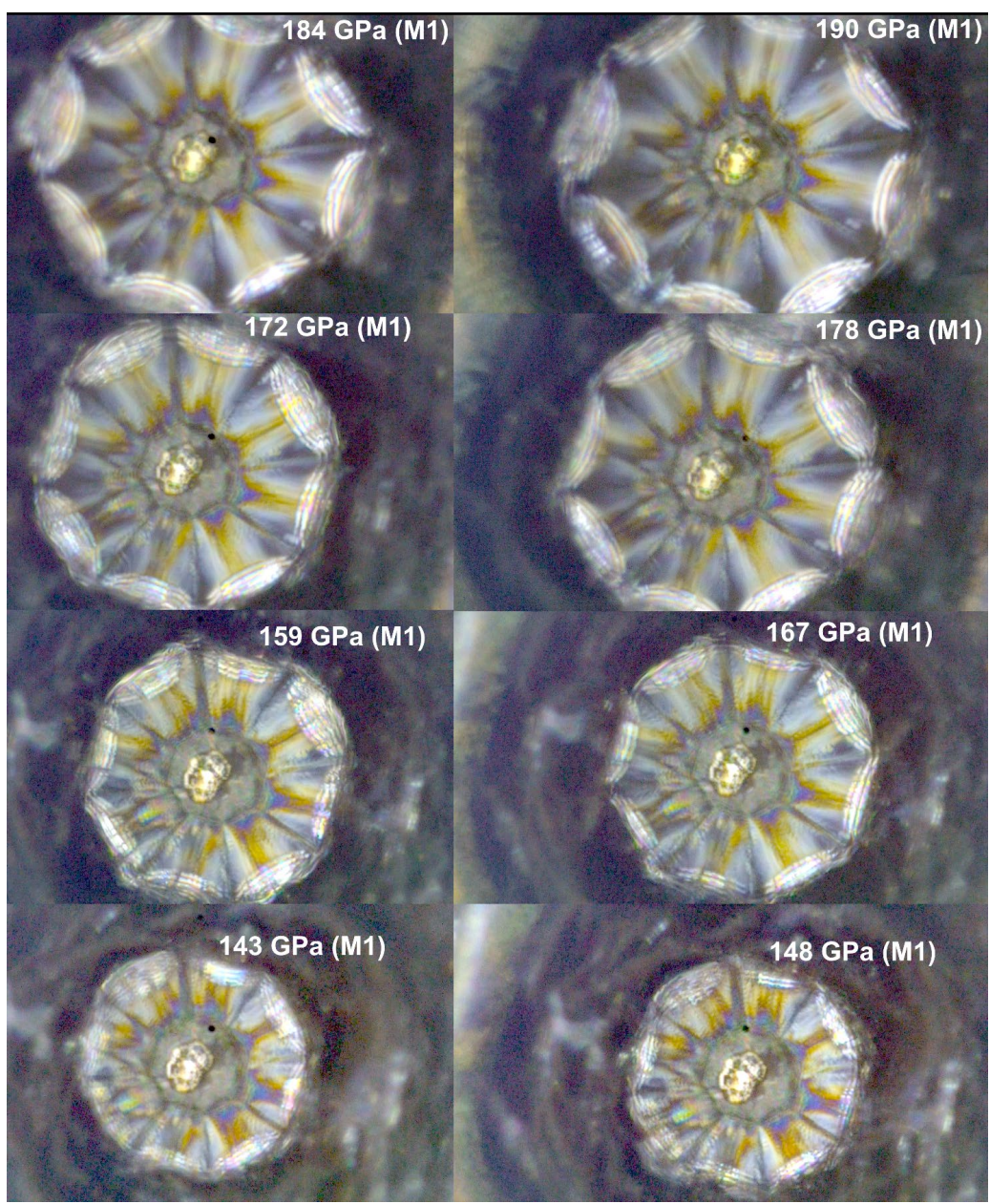

**Figure S5.** Optical photography of the diamond anvil culet of DAC M1 loaded with the La–Nd alloy and AB during the sequential decompression of the cell. The sample retains its characteristic metallic luster at all pressures. The black dot indicates the location where the pressure was measured. Due to the partial cracking of diamond anvils, Raman measurements at exactly the center of the sample were not possible.

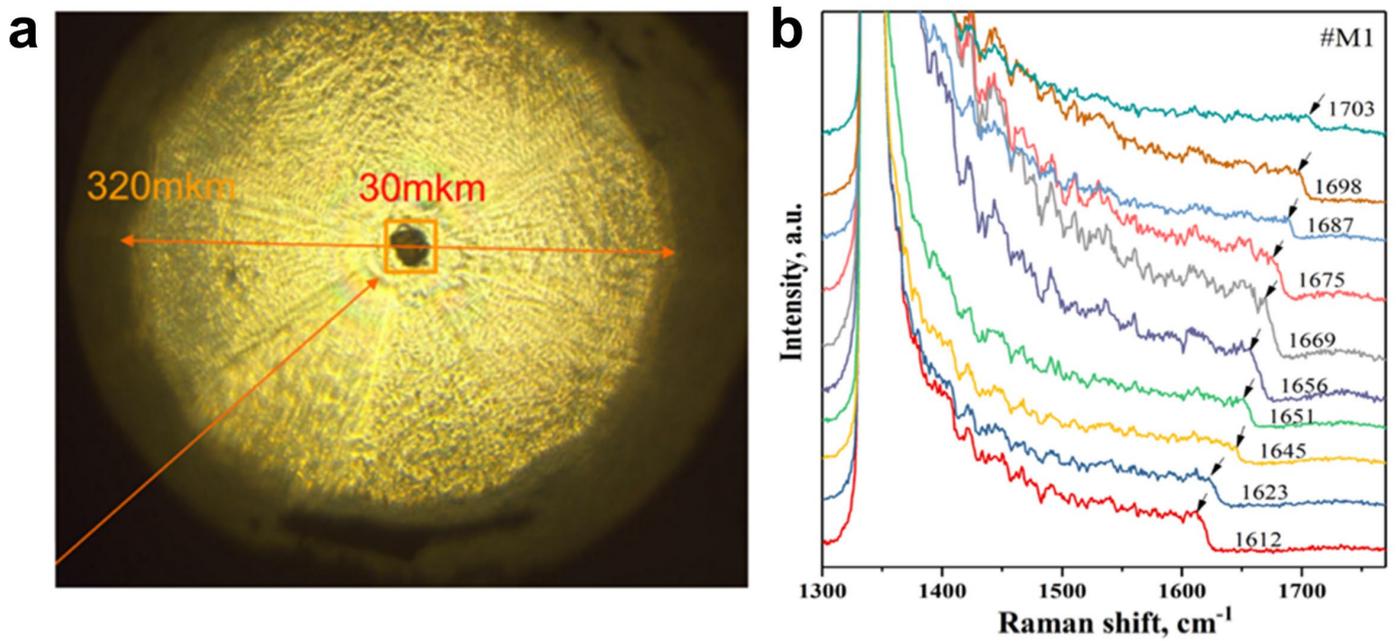

**Figure S6.** (a) Optical image of the sample in DAC M1 at 170 GPa. After the laser heating, the particle expanded and became black. The pressure in DAC M1, measured via the Raman signals of both diamond and hydrogen, was 170 ± 1 GPa. However, during the transportation to SPring-8, the pressure in DAC M1 spontaneously rose to 202 GPa. (b) Raman spectra of DAC M1 during decompression. The numbers indicate the edge frequency of the Raman signal of diamond.

**Table S2.** Experimental lattice parameters and unit cell volume of $Fm\bar{3}m$-La$_{0.91}$Nd$_{0.09}$H$_{10}$ ($Z$ = 4).

| Pressure, GPa | $a$, Å | $V$, Å$^3$ | $V$, Å$^3$/metal atom |
|---|---|---|---|
| 143 | 5.1047 | 133.01 | 33.25 |
| 148 | 5.0896 | 131.84 | 32.95 |
| 159 | 5.0578 | 129.39 | 32.34 |
| 167 | 5.0440 | 128.33 | 32.08 |
| 172 | 5.0250 | 126.88 | 31.72 |
| 178 | 5.0103 | 125.77 | 31.44 |
| 184 | 4.9989 | 124.92 | 31.22 |
| 190 | 4.9827 | 123.71 | 30.92 |
| 197 | 4.9665 | 122.51 | 30.62 |
| 202 | 4.9344 | 120.15 | 30.03 |

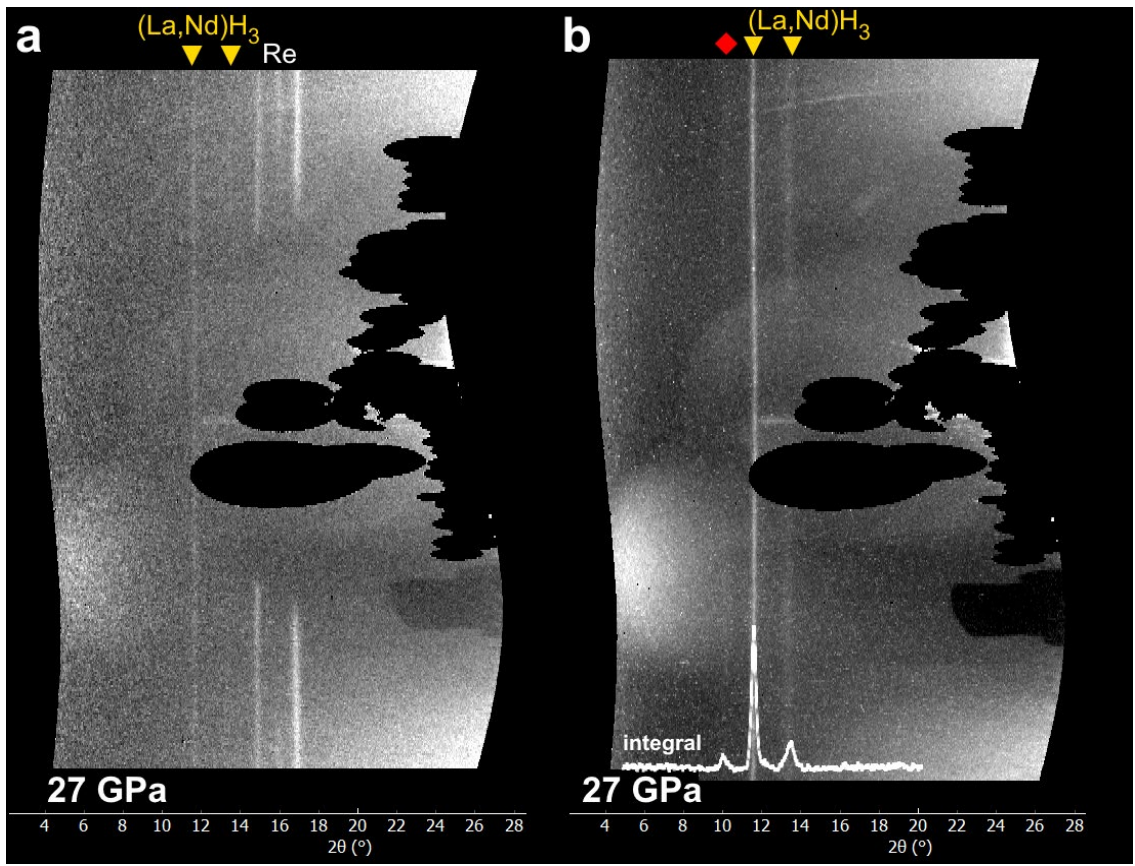

**Figure S7.** Diffraction patterns of lower La-Nd hydrides prepared from the $La_{0.91}Nd_{0.09}$ alloy, studied at the Kurchatov synchrotron radiation source (KISI, $\lambda = 0.62$ Å) in a DAC heated at 35 GPa. After the laser heating, the pressure decreased to 27 GPa. The DAC opening angle does not allow detecting a sufficient number of diffraction rings. However, the expected compound $Fm\bar{3}m$-(La,Nd)$H_3$ ($a = 5.27$ Å, $V = 36.6$ Å$^3$/La, close to Ref. [32]) was apparently present among the reaction products (yellow triangles), along with an unknown hydride (panel b, red diamond). The volume of $LaH_3$[32] and (La,Nd)$H_3$ in the experiment was considerably larger than theoretically predicted ($V_{theory} \sim 33.9$ Å$^3$/La at 27 GPa).

A very similar situation was recently observed by N. Salke et al in a study of the La-Nd-H system with 18 at% of Nd.[33]. The authors experimentally discovered $I4/mmm$-(La,Nd)$H_4$ with the unit cell volume V = 23.43 Å$^3$/f.u. at 170 GPa, whereas theoretical calculations give V(LaH$_4$) = 21.71 Å$^3$/f.u (VASP, PBE GGA) at this pressure. Thus, the experimental equations of state for La and La-Nd hydrides deviate significantly from theoretical predictions.

# Structural information

**Table S3.** Crystal structure of investigated La–Nd–H phases.

| Phase | Pressure, GPa | Lattice parameters | Coordinates | | | |
|---|---|---|---|---|---|---|
| $P\bar{1}$-La$_9$NdH$_{100}$ | 200 | $a$ = 6.0570 Å<br>$b$ = 7.0080 Å<br>$c$ = 7.8340 Å<br>$\alpha$ = 102.71°<br>$\beta$ = 75.31°<br>$\gamma$ = 106.69° | La1(1c) | 0.00000 | 0.50000 | 0.00000 |
| | | | La2(2i) | -0.40029 | -0.40001 | -0.19997 |
| | | | La3(2i) | 0.19950 | -0.30007 | -0.39986 |
| | | | La4(2i) | -0.19974 | -0.19946 | 0.39922 |
| | | | La5(2i) | 0.40097 | -0.09970 | 0.19947 |
| | | | Nd1(1a) | 0.00000 | 0.00000 | 0.00000 |
| | | | H1(2i) | -0.17703 | -0.29505 | -0.01992 |
| | | | H2(2i) | 0.41620 | -0.19774 | -0.21932 |
| | | | H3(2i) | 0.02277 | -0.10001 | -0.42106 |
| | | | H4(2i) | -0.37512 | 0.00205 | 0.37523 |
| | | | H5(2i) | -0.18028 | 0.19955 | -0.01830 |
| | | | H6(2i) | 0.42251 | 0.29986 | -0.22023 |
| | | | H7(2i) | 0.02167 | 0.40009 | -0.41922 |
| | | | H8(2i) | -0.37694 | -0.49969 | 0.37832 |
| | | | H9(2i) | 0.22390 | -0.39877 | 0.17826 |
| | | | H10(2i) | 0.21817 | 0.09973 | 0.17668 |
| | | | H11(2i) | 0.12723 | -0.24496 | -0.12281 |
| | | | H12(2i) | -0.26812 | -0.14715 | -0.32428 |
| | | | H13(2i) | 0.33511 | -0.04803 | 0.47391 |
| | | | H14(2i) | 0.12978 | 0.25080 | -0.12174 |
| | | | H15(2i) | -0.26420 | 0.35166 | -0.32468 |
| | | | H16(2i) | 0.33347 | 0.45217 | 0.47667 |
| | | | H17(2i) | -0.06674 | -0.44659 | 0.27664 |
| | | | H18(2i) | -0.46551 | -0.34663 | 0.07464 |
| | | | H19(2i) | -0.06637 | 0.05418 | 0.27215 |
| | | | H20(2i) | -0.47129 | 0.15088 | 0.07490 |
| | | | H21(2i) | 0.32149 | -0.29412 | -0.02838 |
| | | | H22(2i) | -0.07549 | -0.19225 | -0.22542 |
| | | | H23(2i) | -0.47458 | -0.09571 | -0.42879 |
| | | | H24(2i) | 0.31918 | 0.20224 | -0.02720 |
| | | | H25(2i) | -0.07588 | 0.30154 | -0.22614 |
| | | | H26(2i) | -0.47297 | 0.40300 | -0.42852 |
| | | | H27(2i) | 0.12682 | -0.49631 | 0.37388 |
| | | | H28(2i) | -0.27291 | -0.39473 | 0.17059 |
| | | | H29(2i) | 0.12722 | 0.00387 | 0.36859 |
| | | | H30(2i) | -0.27458 | 0.10680 | 0.16979 |
| | | | H31(2i) | 0.31906 | -0.05327 | -0.02671 |
| | | | H32(2i) | 0.32474 | 0.44437 | -0.02802 |
| | | | H33(2i) | -0.07557 | -0.45375 | -0.22678 |
| | | | H34(2i) | -0.47522 | -0.35537 | -0.42886 |
| | | | H35(2i) | 0.12806 | -0.25432 | 0.37106 |
| | | | H36(2i) | -0.27211 | -0.15084 | 0.16933 |
| | | | H37(2i) | -0.07737 | 0.04407 | -0.22535 |
| | | | H38(2i) | -0.47252 | 0.14408 | -0.42916 |
| | | | H39(2i) | 0.12618 | 0.24436 | 0.37139 |
| | | | H40(2i) | -0.27349 | 0.34553 | 0.17183 |
| | | | H41(2i) | -0.35686 | -0.10099 | -0.05067 |
| | | | H42(2i) | 0.24990 | 0.00210 | -0.25079 |
| | | | H43(2i) | -0.35000 | 0.39998 | -0.05003 |
| | | | H44(2i) | 0.25052 | -0.49980 | -0.24912 |
| | | | H45(2i) | -0.14840 | -0.39941 | -0.45055 |
| | | | H46(2i) | 0.45332 | -0.30082 | 0.34785 |
| | | | H47(2i) | 0.05047 | -0.20106 | 0.15118 |
| | | | H48(2i) | -0.14656 | 0.09809 | -0.45296 |
| | | | H49(2i) | 0.44999 | 0.19945 | 0.34988 |
| | | | H50(2i) | 0.05070 | 0.30712 | 0.15355 |
| $C2/m$-La$_4$NdH$_{50}$ | 200 | $a$ = 11.5686 Å<br>$b$ = 3.5030 Å<br>$c$ = 8.5788 Å<br>α = γ = 90°<br>β = 119.52° | La1(4i) | 0.09993 | 0.00000 | 0.39863 |
| | | | La2(4i) | -0.30034 | 0.00000 | -0.19963 |
| | | | Nd1(2b) | 0.00000 | 0.50000 | 0.00000 |
| | | | H1(4i) | 0.19592 | 0.00000 | -0.07422 |
| | | | H2(8j) | -0.05106 | 0.24062 | 0.16851 |
| | | | H3(4i) | -0.49938 | 0.00000 | 0.37497 |
| | | | H4(8j) | 0.14657 | 0.24172 | -0.02672 |
| | | | H5(4i) | 0.09965 | 0.00000 | -0.22083 |
| | | | H6(8j) | -0.35035 | 0.24185 | 0.22260 |
| | | | H7(8j) | -0.25009 | 0.24166 | 0.37284 |
| | | | H8(4i) | 0.10315 | 0.00000 | 0.15372 |
| | | | H9(4i) | -0.00414 | 0.00000 | 0.11998 |
| | | | H10(4i) | -0.30141 | 0.00000 | 0.41680 |
| | | | H11(4i) | -0.30025 | 0.00000 | -0.45020 |
| | | | H12(4i) | -0.09765 | 0.00000 | -0.01894 |
| | | | H13(4i) | 0.29650 | 0.00000 | -0.05100 |
| | | | H14(4i) | -0.29986 | 0.00000 | 0.17891 |
| | | | H15(4i) | -0.39554 | 0.00000 | 0.27426 |

| | | | H16(4i) | 0.49993 | 0.00000 | 0.24924 |
| --- | --- | --- | --- | --- | --- | --- |
| | | | H17(4i) | -0.20500 | 0.00000 | 0.32382 |
| | | | H18(4i) | -0.40504 | 0.00000 | -0.47634 |
| | | | H19(8j) | 0.45187 | 0.24057 | 0.43007 |
| | | | H20(4i) | 0.09998 | 0.00000 | -0.34868 |
| $I4/m$-La$_4$NdH$_{50}$ | 200 | $a = b = 7.8230$ Å<br>$c = 4.9480$ Å<br>$\alpha = \gamma = \beta = 90°$ | La1(8h) | -0.29996 | 0.10006 | 0.00000 |
| | | | Nd1(2b) | 0.00000 | 0.00000 | 0.50000 |
| | | | H1(16i) | 0.20427 | 0.10363 | 0.24044 |
| | | | H2(16i) | -0.09489 | -0.04643 | 0.12382 |
| | | | H3(16i) | 0.30049 | -0.35420 | 0.11776 |
| | | | H4(16i) | 0.00392 | 0.24839 | 0.38489 |
| | | | H5(16i) | 0.05007 | 0.39445 | 0.37976 |
| | | | H6(16i) | 0.34830 | -0.19698 | 0.37505 |
| | | | H7(4d) | 0.00000 | 0.50000 | 0.25000 |
| $C2/m$-La$_2$Nd$_3$H$_{50}$ | 200 | $a = 11.3610$ Å<br>$b = 3.4319$ Å<br>$c = 10.2580$ Å<br>$\alpha = \beta = 90°$<br>$\gamma = 134.89°$ | La1(4i) | 0.29824 | 0.00000 | 0.39872 |
| | | | Nd1(4i) | 0.09967 | 0.00000 | -0.20161 |
| | | | Nd2(2b) | 0.00000 | 0.50000 | 0.00000 |
| | | | H1(8j) | -0.27884 | 0.25411 | 0.16673 |
| | | | H2(8j) | 0.32769 | 0.25916 | -0.02180 |
| | | | H3(8j) | 0.06581 | 0.25901 | 0.21820 |
| | | | H4(8j) | 0.11458 | 0.25562 | 0.36487 |
| | | | H5(8j) | 0.47408 | 0.25628 | 0.42393 |
| | | | H6(4i) | -0.27219 | 0.00000 | -0.07338 |
| | | | H7(4i) | -0.31673 | 0.00000 | -0.21416 |
| | | | H8(4i) | 0.11924 | 0.00000 | 0.11641 |
| | | | H9(4i) | -0.34283 | 0.00000 | -0.04387 |
| | | | H10(4i) | 0.47150 | 0.00000 | 0.17232 |
| | | | H11(4i) | -0.48123 | 0.00000 | 0.31542 |
| | | | H12(4i) | -0.44825 | 0.00000 | -0.34889 |
| | | | H13(4i) | 0.04816 | 0.00000 | 0.14798 |
| | | | H14(4i) | 0.08331 | 0.00000 | -0.02127 |
| | | | H15(4i) | -0.12697 | 0.00000 | 0.37288 |
| | | | H16(4i) | -0.27939 | 0.00000 | 0.42119 |
| | | | H17(4i) | -0.14408 | 0.00000 | -0.44526 |
| | | | H18(4i) | -0.33601 | 0.00000 | 0.26525 |
| | | | H19(4i) | -0.25504 | 0.00000 | 0.24336 |
| | | | H20(4i) | -0.06957 | 0.00000 | -0.47405 |
| $C2/m$-La$_3$Nd$_2$H$_{50}$ | 200 | $a = 11.4142$ Å | La1(4i) | 0.09848 | 0.00000 | 0.30011 |
| | | $b = 3.4426$ Å<br>$c = 10.3090$ Å<br>$\alpha = \beta = 90°$<br>$\gamma = 134.81°$ | La2(2d) | 0.00000 | 0.50000 | 0.50000 |
| | | | Nd1(4i) | -0.30030 | 0.00000 | 0.10061 |
| | | | H1(8j) | -0.47553 | 0.25602 | 0.07084 |
| | | | H2(8j) | 0.08654 | 0.25468 | -0.26391 |
| | | | H3(8j) | 0.13435 | 0.25649 | -0.11817 |
| | | | H4(8j) | -0.27110 | 0.25836 | -0.31965 |
| | | | H5(8j) | -0.32228 | 0.25642 | -0.47002 |
| | | | H6(4i) | -0.46559 | 0.00000 | -0.16645 |
| | | | H7(4i) | 0.07715 | 0.00000 | -0.02138 |
| | | | H8(4i) | 0.48536 | 0.00000 | -0.31664 |
| | | | H9(4i) | 0.11703 | 0.00000 | 0.11687 |
| | | | H10(4i) | -0.31626 | 0.00000 | -0.21338 |
| | | | H11(4i) | 0.45532 | 0.00000 | -0.14567 |
| | | | H12(4i) | 0.14864 | 0.00000 | -0.04923 |
| | | | H13(4i) | 0.27692 | 0.00000 | 0.07650 |
| | | | H14(4i) | 0.32835 | 0.00000 | -0.26874 |
| | | | H15(4i) | 0.27379 | 0.00000 | -0.42181 |
| | | | H16(4i) | 0.34998 | 0.00000 | -0.45169 |
| | | | H17(4i) | 0.24919 | 0.00000 | 0.25108 |
| | | | H18(4i) | -0.07453 | 0.00000 | -0.47681 |
| | | | H19(4i) | -0.05560 | 0.00000 | 0.34681 |
| | | | H20(4i) | 0.13456 | 0.00000 | -0.36809 |
| $P4/mmm$-LaNdH$_{20}$ | 200 | $a = b = 3.4340$ Å<br>$c = 4.8310$ Å<br>$\alpha = \beta = \gamma = 90°$ | La1(1a) | 0.00000 | 0.00000 | 0.00000 |
| | | | Nd1(1d) | 0.50000 | 0.50000 | 0.50000 |
| | | | H1(8t) | 0.25847 | 0.50000 | 0.12716 |
| | | | H2(8s) | 0.24546 | 0.00000 | 0.38211 |
| | | | H3(4i) | 0.00000 | 0.50000 | -0.25233 |
| $Immm$-La$_3$H$_8$ | 200 | $a = 2.69300$ Å<br>$b = 15.3060$ Å<br>$c = 2.6450$ Å<br>$\alpha = \beta = \gamma = 90°$ | La1(4g) | 0.00000 | 0.34195 | 0.00000 |
| | | | La2(2c) | 0.50000 | 0.50000 | 0.00000 |
| | | | H1(4g) | 0.00000 | 0.10020 | 0.00000 |
| | | | H2(4g) | 0.00000 | 0.20742 | 0.00000 |
| | | | H3(4h) | 0.00000 | -0.44821 | 0.50000 |
| | | | H4(4h) | 0.00000 | -0.25057 | 0.50000 |

# Thermodynamic stability of La–Nd hydrides

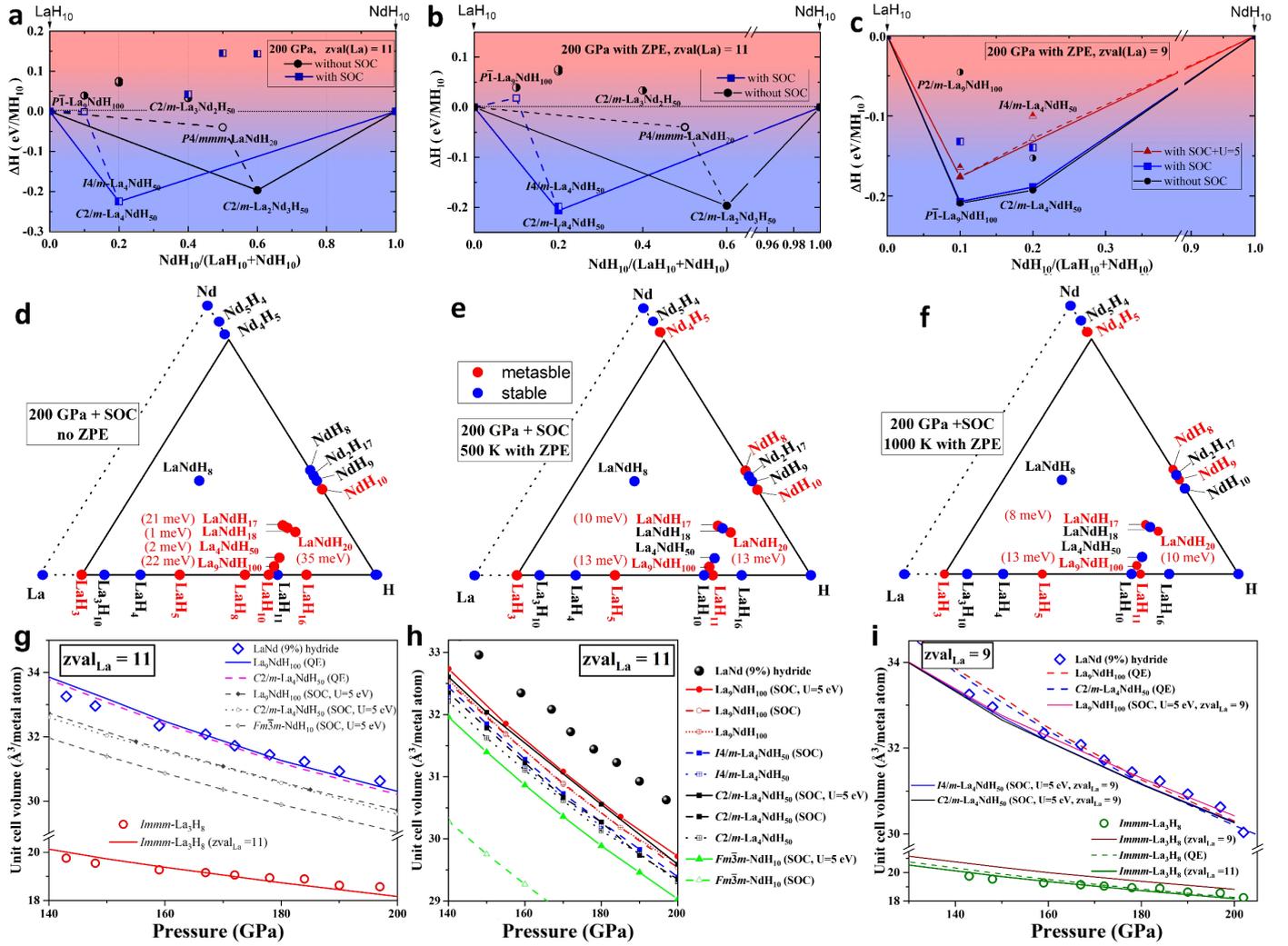

**Figure S8**. Convex hulls of La–Nd hydrides at 200 GPa. (a–c) Convex hulls of La–Nd hydrides with respect to decomposition to $LaH_{10}$ and $NdH_{10}$: (a) calculated using VASP with the La pseudopotential with 11 outer shell electrons, with and without the spin–orbit coupling (SOC); (b) additionally, the ZPE is included; (c) calculated using the La pseudopotential with 9 outer shell electrons at 200 GPa, including the ZPE, with and without the SOC and the Hubbard term $U - J = 5$ eV (Nd, as was done in the Ref. [34]). (d–f) Convex hulls of the La–Nd–H system (stable and metastable phases are shown in blue and red, respectively): (d) with the SOC at 0 K, without the ZPE effect; (e) with the SOC and ZPE at 500 K; (f) with the SOC and ZPE at 1000 K. (g–i) Equations of state (EOS) of La–Nd hydrides calculated using the VASP pseudopotentials with 9 and 11 outer shell electrons, with and without the SOC and $U - J$. Experimentally obtained data is shown by open circles, diamonds and filled spheres. The results obtained using Quantum ESPRESSO are marked QE. In all cases, the PBE GGA functional was used.

Ab initio calculations with 110 atoms are a big challenge. To simplify the computational problem when describing La–Nd–H system, we considered the following fixed compositions (the main models): pseudocubic $La_9NdH_{100}$ (10 at% of Nd), $La_4NdH_{50}$ (20 at% of Nd), and $LaNdH_{20}$ (50 at% of Nd). The DFT+U approach with $U - J = 5$ eV for Nd atoms better explains the experimental equation of states (EOS, Figure S8g–i). When the La pseudopotential with 11 outer shell electrons is used, the unit cell volumes calculated using VASP[17,18,35] are lower than the experimental ones by 1 Å³ for both $La_9NdH_{100}$ and $La_4NdH_{50}$, with and without the SOC and $U - J = 5$ eV for Nd (Figure S8g–h). The similar problem of the $LaH_{10}$ equation of state is that different series of experimental measurements and theoretical calculations show that the composition of lanthanum superhydride may vary within $LaH_{10\pm0.5}$ (or between $La_2H_{19}$ and $La_2H_{21}$) without changing the

S15

superconducting properties.

Calculations of the EOS using the PBE La pseudopotentials in Quantum ESPRESSO[24,25] yield much better agreement with experiment for both La$_4$NdH$_{50}$ and La$_9$NdH$_{100}$ (Figure S8g–i). The calculated equation of state of *Immm*-La$_3$H$_8$ (LaH$_{2.66}$), whose formation may be explained by the reaction (La$_9$Nd)H$_{10}$ → (La$_4$Nd)H$_{10}$ + La$_3$H$_8$, is also in close agreement with the XRD results.

**Table S4.** Temperature dependence of the Gibbs free energy of formation ($G_{form}$, eV/atom), computed with the ZPE and SOC for various La–Nd–H phases at 200 GPa. "2000 K", "1000 K" and "500 K" marks in the header denote the temperature used in the calculations.

| Phase | $E_{form}$, eV/atom | ZPE, eV/atom | $E_{form}$+ZPE, eV/atom | $G_{form}$ (2000 K), eV/atom | $G_{form}$ (1000 K), eV/atom | $G_{form}$ (500 K), eV/atom |
|---|---|---|---|---|---|---|
| $Fm\bar{3}m$-La | 0 | 0.0451 | 0 | 0 | 0 | 0 |
| *Immm*-La$_3$H$_8$ | -0.7006 | 0.2204 | -0.6985 | -0.6112 | -0.6665 | -0.6912 |
| *Cmcm*-LaH$_3$ | -0.6374 | 0.2303 | -0.6308 | -0.5423 | -0.5970 | -0.6222 |
| *Cmmm*-La$_3$H$_{10}$ | -0.6925 | 0.2368 | -0.6839 | -0.5856 | -0.6457 | -0.6741 |
| *I*4/*mmm*-LaH$_4$ | -0.6502 | 0.2389 | -0.6468 | -0.5451 | -0.6059 | -0.6356 |
| $P\bar{1}$-LaH$_5$ | -0.5704 | 0.2412 | -0.5726 | -0.4868 | -0.5378 | -0.5634 |
| $Fm\bar{3}m$-LaH$_{10}$ | -0.3880 | 0.2227 | -0.4268 | -0.3639 | -0.3981 | -0.4179 |
| *P*4/*nmm*-LaH$_{11}$ | -0.3801 | 0.2644 | -0.3790 | -0.2887 | -0.3415 | -0.3681 |
| *P*6/*mmm*-LaH$_{16}$ | -0.2655 | 0.2256 | -0.3090 | -0.1815 | -0.2663 | -0.2981 |
| *C*2/*c*-H | 0 | 0.2831 | 0 | 0 | 0 | 0 |
| *C*2/*m*-Nd | 0 | 0.0432 | 0 | 0 | 0 | 0 |
| $P3m1$-Nd$_5$H$_4$ | -0.3239 | 0.0464 | -0.4273 | -0.2879 | -0.3868 | -0.4161 |
| *C*2/*m*-Nd$_4$H$_5$ | -0.3507 | 0.2061 | -0.3211 | -0.2920 | -0.3001 | -0.3132 |
| $Fm\bar{3}m$-NdH$_8$ | -0.3753 | 0.2410 | -0.3907 | -0.3067 | -0.3531 | -0.3786 |
| $R3m$-Nd$_2$H$_{17}$ | -0.3675 | 0.2314 | -0.3940 | -0.3175 | -0.3595 | -0.3828 |
| $F\bar{4}3m$-NdH$_9$ | -0.3491 | 0.2331 | -0.3751 | -0.2969 | -0.3398 | -0.3637 |
| $Fm\bar{3}m$-NdH$_{10}$ | -0.2989 | 0.2236 | -0.3366 | -0.3000 | -0.3148 | -0.3284 |
| *C*2/*m*-LaNdH$_8$ | -0.5449 | 0.2186 | -0.5616 | -0.4339 | -0.5214 | -0.5591 |
| $P3m1$-LaNdH$_{17}$ | -0.3785 | 0.2239 | -0.4125 | -0.3486 | -0.3873 | -0.4088 |
| $R\bar{3}m$-LaNdH$_{18}$ | -0.3877 | 0.2314 | -0.4156 | -0.3475 | -0.3879 | -0.4108 |
| *P*4/*mmm*-LaNdH$_{20}$ | -0.3306 | 0.2186 | -0.3734 | -0.3205 | -0.3524 | -0.3702 |
| *P*1-La$_9$NdH$_{100}$ | -0.3702 | 0.2233 | -0.4083 | -0.3509 | -0.3813 | -0.3996 |
| *C*/2*m*-La$_4$NdH$_{50}$ | -0.3830 | 0.2272 | -0.4173 | -0.3585 | -0.3905 | -0.4091 |
| *I*4/*m*-La$_4$NdH$_{50}$ | -0.3830 | 0.2314 | -0.4130 | -0.3523 | -0.3855 | -0.4047 |

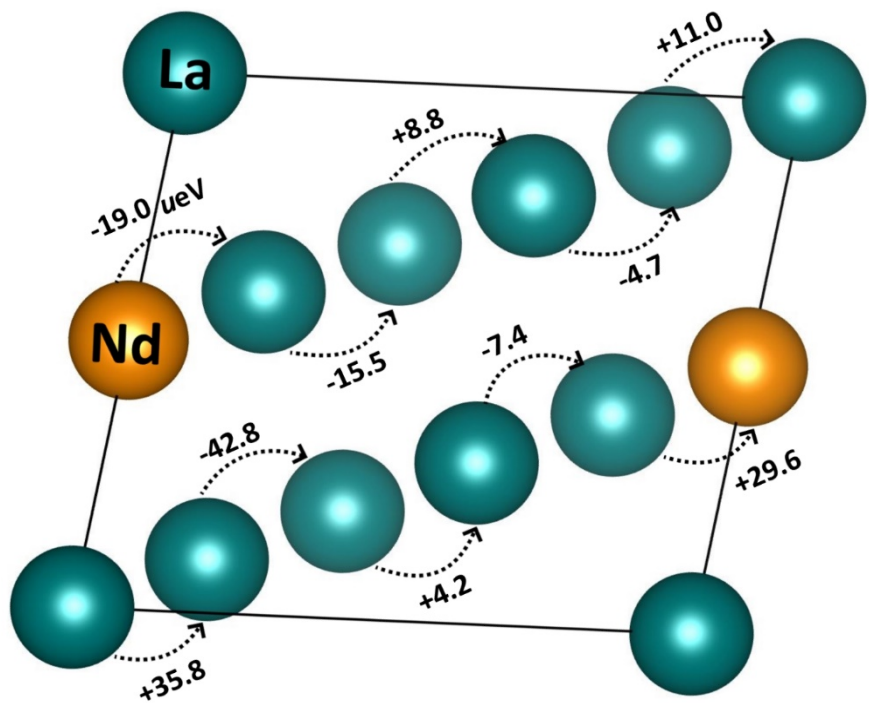

**Figure S9**. Migration (indicated by arrows) of neodymium atoms in the LaH$_{10}$ lattice. The captions to the arrows are the formation energy differences ΔH in µeV/atom berween the corresponding $P$1-La$_9$NdH$_{100}$ modifications. For simplicity, hydrogen is not shown. Such low energy barriers between structures with different distributions of Nd atoms indicate that in the real structure the Nd atoms will be randomly distributed in the La sublattice of LaH$_{10}$.

# Electron, phonon band structures and superconductivity of La–Nd hydrides

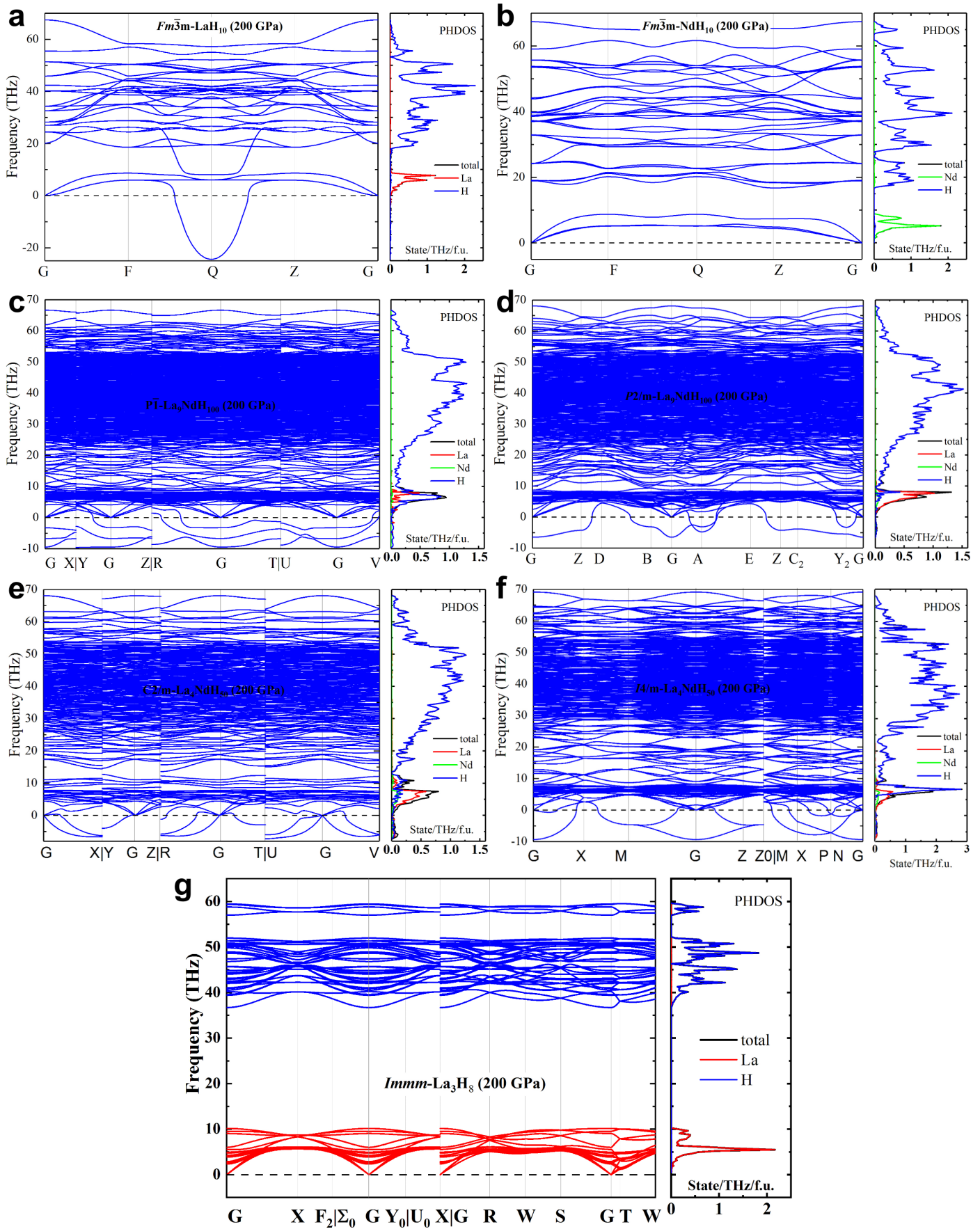

**Figure S10.** Harmonic phonon band structure and density of states for various La–Nd hydrides at 200 GPa. Almost all models of ternary La–Nd hydrides have imaginary phonon modes in the harmonic approximation.

S18

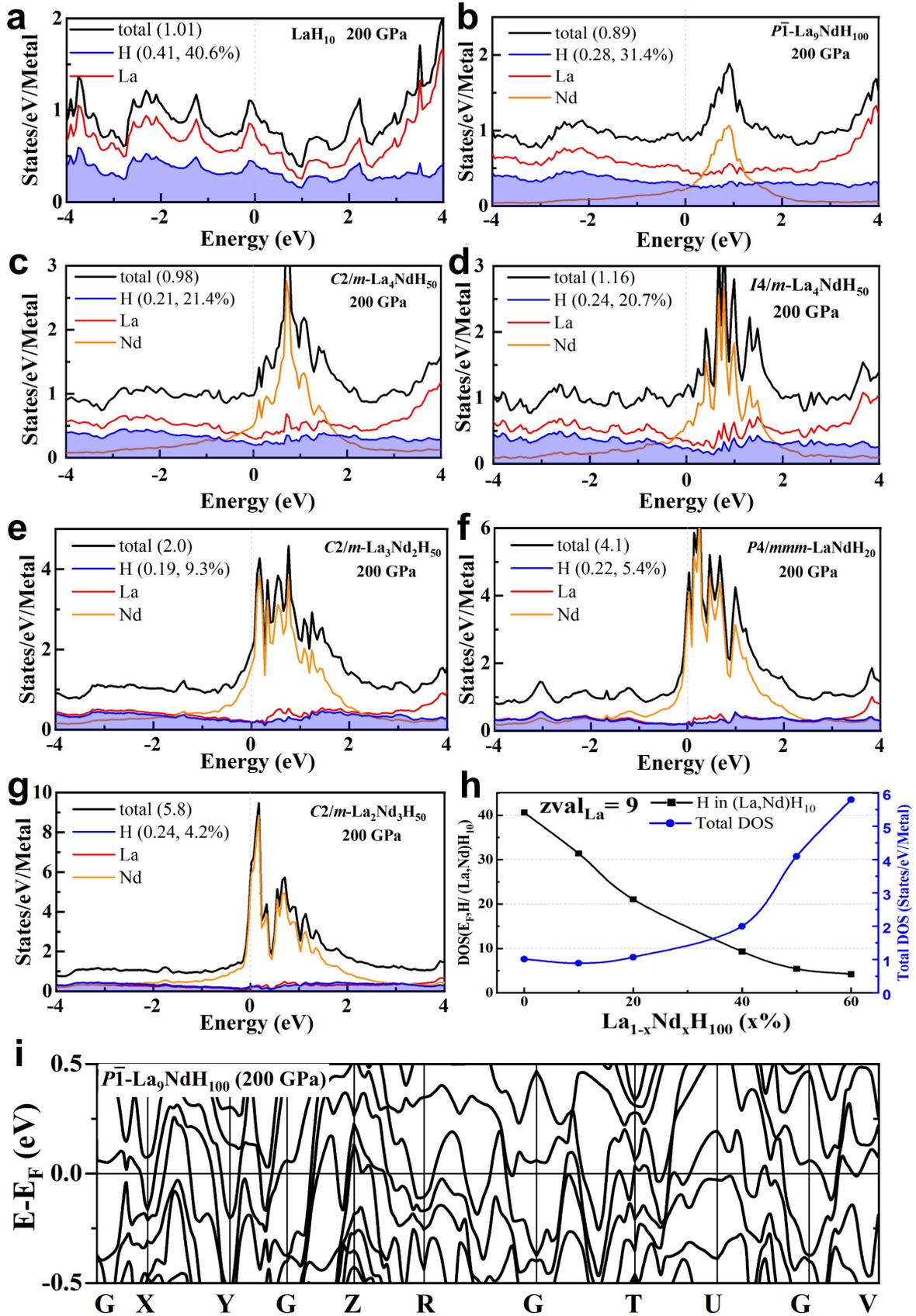

**Figure S11.** (a–i) Electronic density of states for various La–Nd hydrides at 200 GPa. With an increase in the Nd content, the density of states at the Fermi level increases because of the $d, f$ electrons of neodymium, whereas the relative contribution of hydrogen decreases. (h) Dependence of the calculated electronic DOS at the Fermi level, total and projected on H (in percent), on the Nd content. (i) Electronic band structure near the Fermi energy of $P1$-La$_9$NdH$_{100}$ at 200 GPa.

The comparison of the density of states at the Fermi level DOS($E_F$) for La$_{1-x}$Nd$_x$H$_{10}$ (Figure S11) shows that Nd atoms, which have the f electrons, increase the total DOS($E_F$), but the contribution of the H sublattice decreases correspondingly. As a rule, high-temperature hydride superconductors have a relatively high contribution of hydrogen to the density of electronic states[36,37]. In the limit of zero H-contribution to DOS, we arrive at superconductivity in pure metals, where $T_C$ does not exceed 30 K even at high pressures. Here, an increase in the Nd concentration leads to a decrease in the relative contribution of hydrogen to the total density of states, which qualitatively corresponds to a decrease in the critical temperature of superconductivity of $(La,Nd)H_{10}$.

On the other hand, $T_C$ is generically expected to rise when an increase in the number of superconducting carriers leads to filling of additional conduction bands: the density of states increases, and the superconducting coupling increases accordingly, with important effects on $T_C$. However, in our case this idea obviously does not work, primarily, due to the disordered arrangement of Nd atoms. Moreover, an increase in the density of electronic states at the Fermi level in hydrides of heavy lanthanides stabilizes the macroscopic magnetic ordering, which is not good for superconductivity.

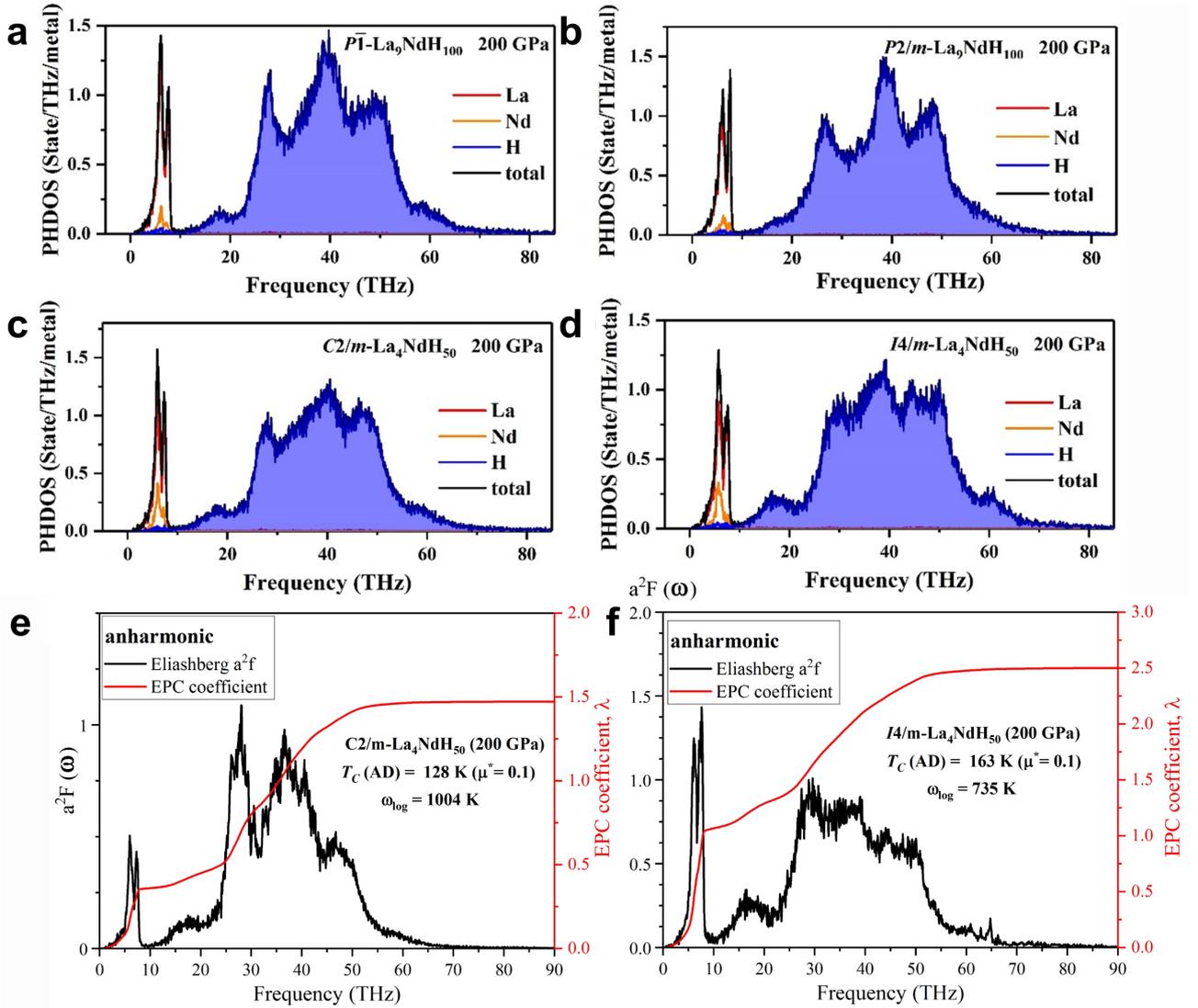

**Figure S12.** (a–d) Anharmonic phonon density of states at 200 GPa and 300 K for $P\bar{1}$ and $P2/m$ phases of $La_9NdH_{100}$ and $C2/m$ and $I4/m$ phases of $La_4NdH_{50}$. The calculations show that the model ternary La–Nd hydrides are dynamically stable in the anharmonic approximation and can be used for a theoretical description of compounds synthesized in the experiment. (e, f) Anharmonic Eliashberg functions and integrated EPC parameters λ for pseudocubic $C2/m$- and $I4/m$-$La_4NdH_{50}$ at 200 GPa. The calculations were performed using the same methodology that was used in the works on europium hydrides[38] and La-Y hydrides[22].

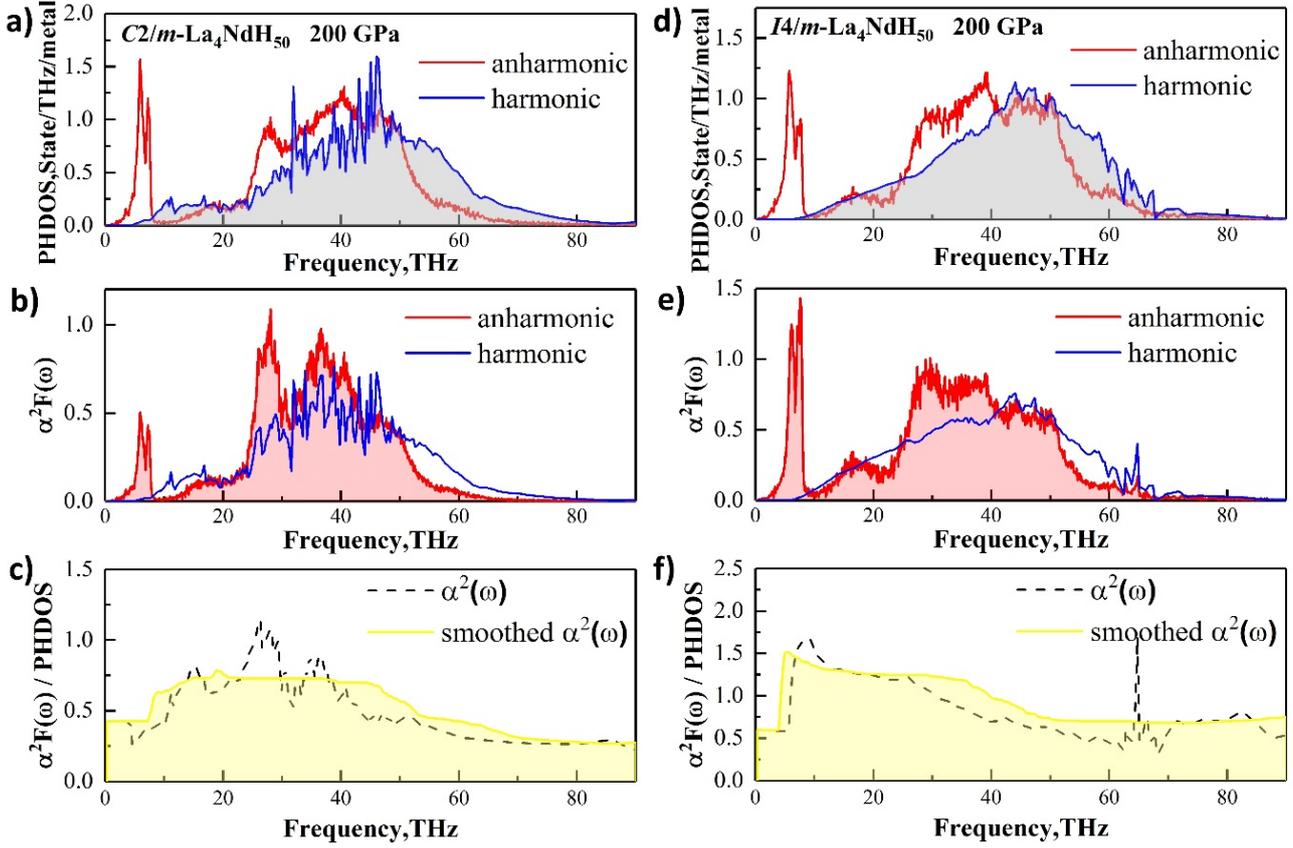

**Figure S13.** Phonon density of states and the Eliashberg function obtained in the harmonic and anharmonic calculations. (a, d) Phonon density of states (PHDOS) for $C2/m$- and $I4/m$-La$_4$NdH$_{50}$ at 200 GPa. (b, e) Eliashberg functions for $C2/m$- and $I4/m$-La$_4$NdH$_{50}$. (c, f) Contribution of the electron band structure to the Eliashberg functions. The calculations were performed using the same methodology that was used in the works on europium hydrides[38] and La–Y hydrides[22].

Complete suppression of superconductivity due to scattering by paramagnetic impurities cannot be predicted using the standard DFT calculations in Quantum ESPRESSO.[24,25] To compare the experimental results with theoretical predictions, we calculated the superconducting properties of the model La–Nd–H phases: pseudocubic $C2/m$-La$_4$NdH$_{50}$, $I4/m$-La$_4$NdH$_{50}$, $P4/mmm$-LaNdH$_{20}$, $C2/m$-La$_2$Nd$_3$H$_{50}$, and $Fm\bar{3}m$-NdH$_{10}$. All these phases are built from $Fm\bar{3}m$-LaH$_{10}$ by partially replacing the La atoms with Nd. The results of the calculations of the electron–phonon coupling parameters using the LDA Perdew–Zunger (PZ) exchange–correlation functional are in qualitative agreement with the experimental observations: $T_C$ of La$_{1-x}$Nd$_x$H$_{10}$ decreases as the Nd concentration rises (Figure S15). Using the PBE pseudopotentials for La and Nd leads to overestimation of $T_C$ for all La$_{1-x}$Nd$_x$H$_{10}$ compositions including NdH$_{10}$ (using the Allen–Dynes (AD) formula with $\mu^* = 0.1$ results in $T_C = 266$ K, which is far from the experimental data [34] (Figure S14)).

Comparing the two crystal modifications of La$_4$NdH$_{50}$ ($C2/m$ and $I4/m$, Figure S12, S14-15) which have a very similar XRD pattern, it should be noted that there are reasons, except the spin flip scattering discussed earlier, why Nd may suppress superconductivity in LaH$_{10}$. Doping by Nd slightly distorts the H sublattice of LaH$_{10}$ and reduces its high symmetry. Even though Anderson's theorem is well satisfied in slightly disordered systems, we should remember that the presence of a strong disorder, as indicated by wide experimental superconducting transitions ($\Delta T_C > 30$ K), is by itself an effective pair-breaking mechanism that suppresses superconductivity.[39]

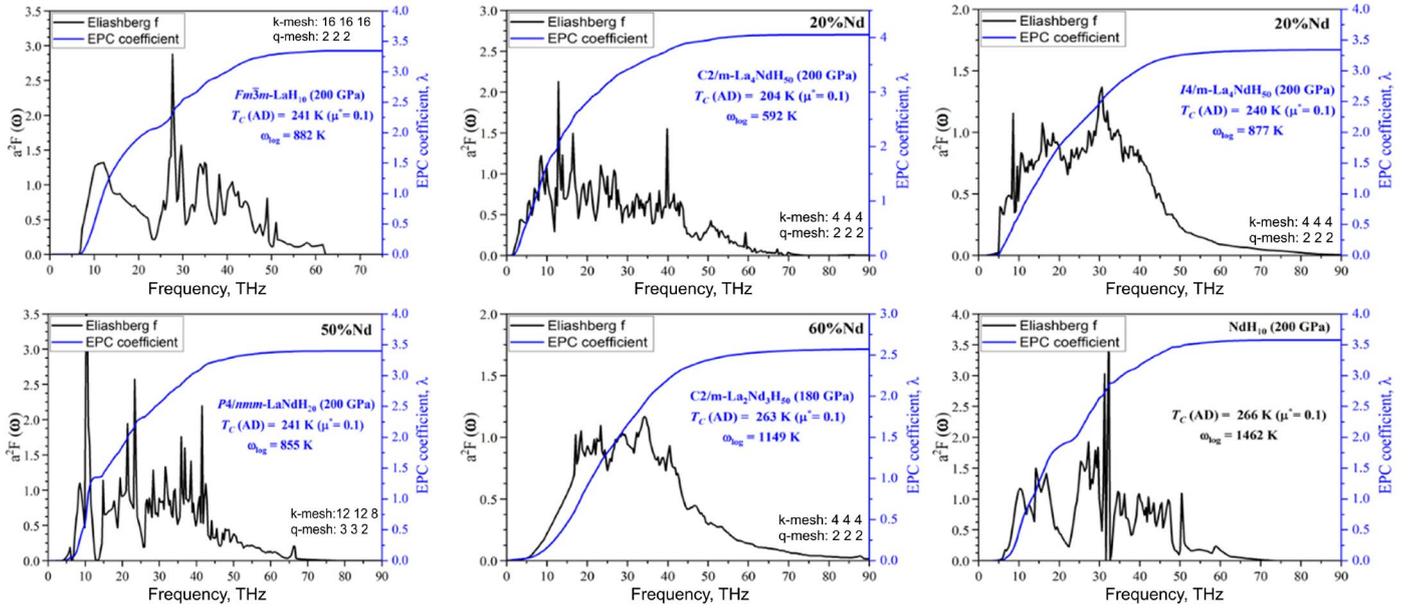

**Figure S14.** Eliashberg functions and integrated λ(ω) of the La–Nd polyhydrides at 200 GPa calculated using Quantum ESPRESSO (QE) with the PBE GGA pseudopotentials and various *q*- and *k*-meshes. Neodymium does not suppress superconductivity in this series of QE calculations.

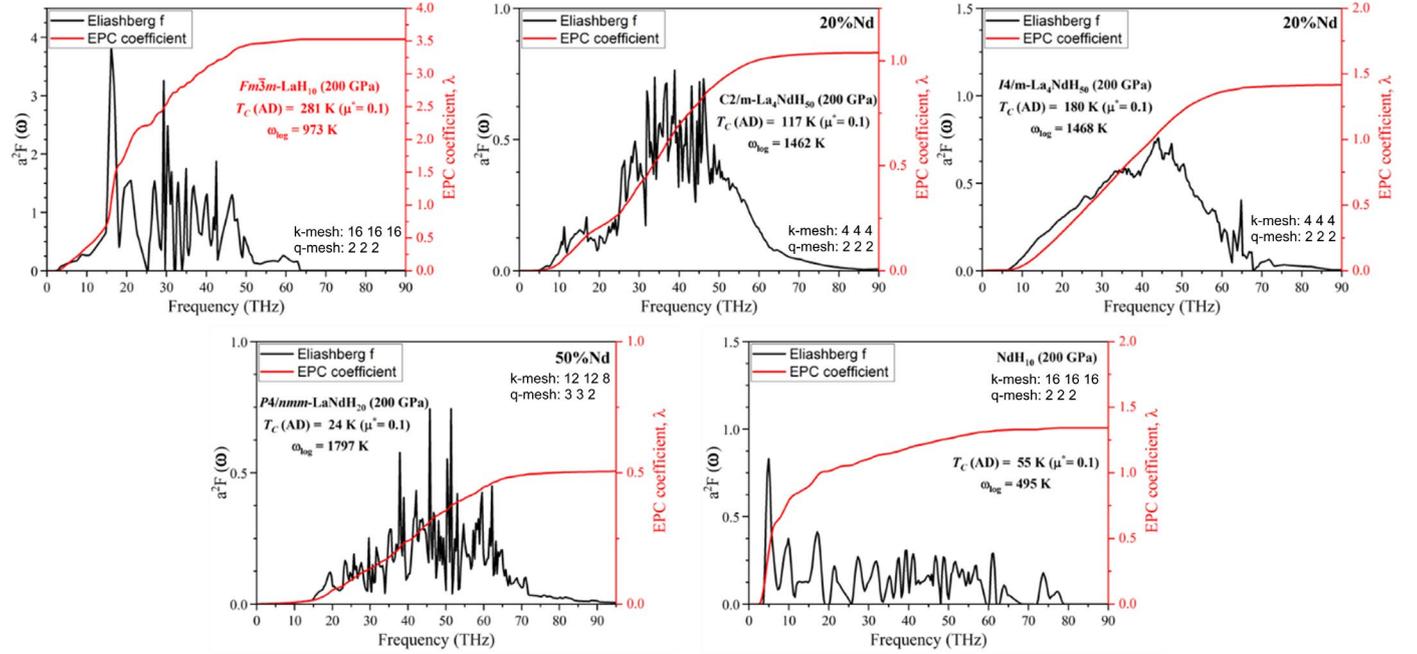

**Figure S15.** Eliashberg functions and integrated λ(ω) of the La–Nd polyhydrides at 200 GPa calculated using Quantum ESPRESSO with the norm-conserving Perdew-Zunger LDA pseudopotentials and various *q*- and *k*-meshes. These calculations indicate that Nd suppresses superconductivity.

**Table S5.** Superconducting parameters for Nd-doped $LaH_{10}$ calculated at 200 GPa using the norm-conserving Perdew-Zunger LDA pseudopotentials.

| Nd doping, % | $T_C$, K | λ | $\omega_{\log}$, K |
| --- | --- | --- | --- |
| 0 | 281 | 3.54 | 973 |
| 20 (*C2/m*) | 117 | 1.04 | 1462 |
| 20 (*I4/m*) | 180 | 1.43 | 1468 |
| 50 | 24 | 0.51 | 1797 |
| 100 | 55 | 1.31 | 500 |

The number of atoms in the unit cell of pseudocubic $P\bar{1}$-La$_9$NdH$_{100}$ is too large for a direct calculation of $T_C$, therefore we constructed the virtual crystal approximation (VCA) appended with the La–Nd norm-conserving Perdew-Zunger pseudopotentials. For this compound, the calculations within the VCA at 200 GPa give $T_C$ = 150 K, close to the experimental value (Figure S16). However, none of the variants of mixed La–Nd pseudopotentials tested in this work made it possible to describe, even qualitatively, the dependence of $T_C$ on the Nd concentration in the entire range.

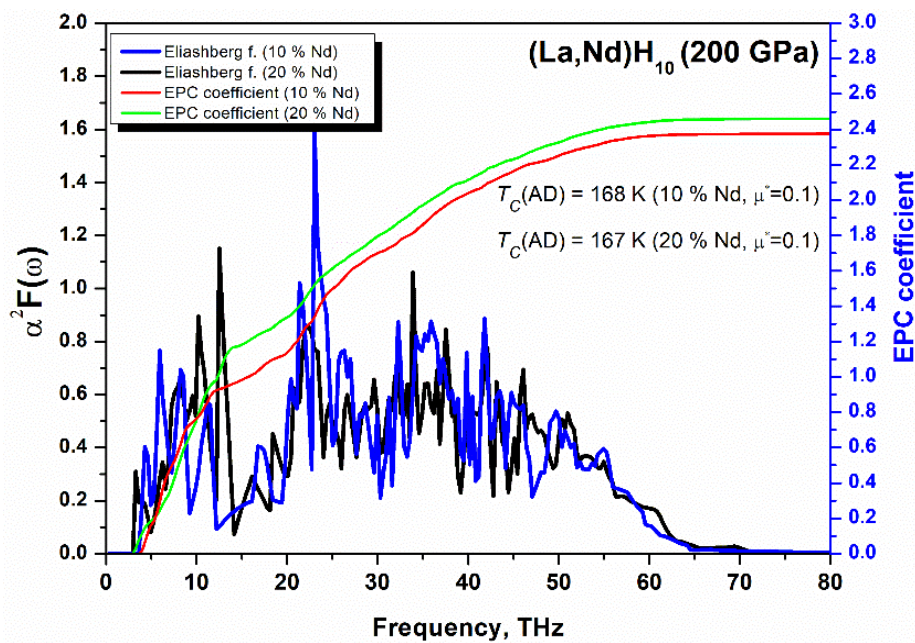

**Figure S16.** Eliashberg functions of (La,Nd)H$_{10}$ with 10 at% and 20 at% of Nd calculated using the VCA PBE pseudopotentials. Almost identical superconducting properties for different compounds indicate the inefficiency of the proposed mixed La–Nd pseudopotentials.

# Pulse magnetic and transport measurements in DACs

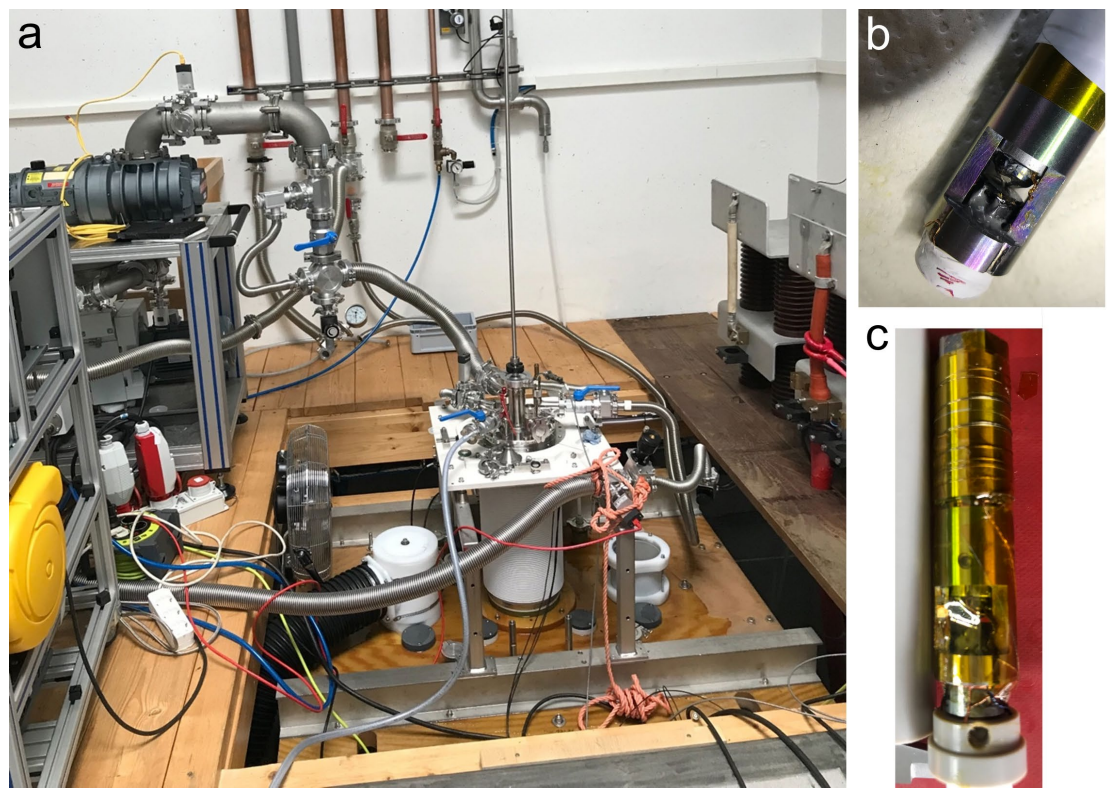

**Figure S17.** Pulsed magnetic measurements. (a) 72 T pulsed magnet system and (b, c) Ni–Cr–Al diamond anvil cell E1 used for determining the magnetic phase diagram of $La_{0.91}Nd_{0.09}H_{10}$ at 170 GPa. Diameter of the diamond anvil cell is 15.1–15.3 mm.

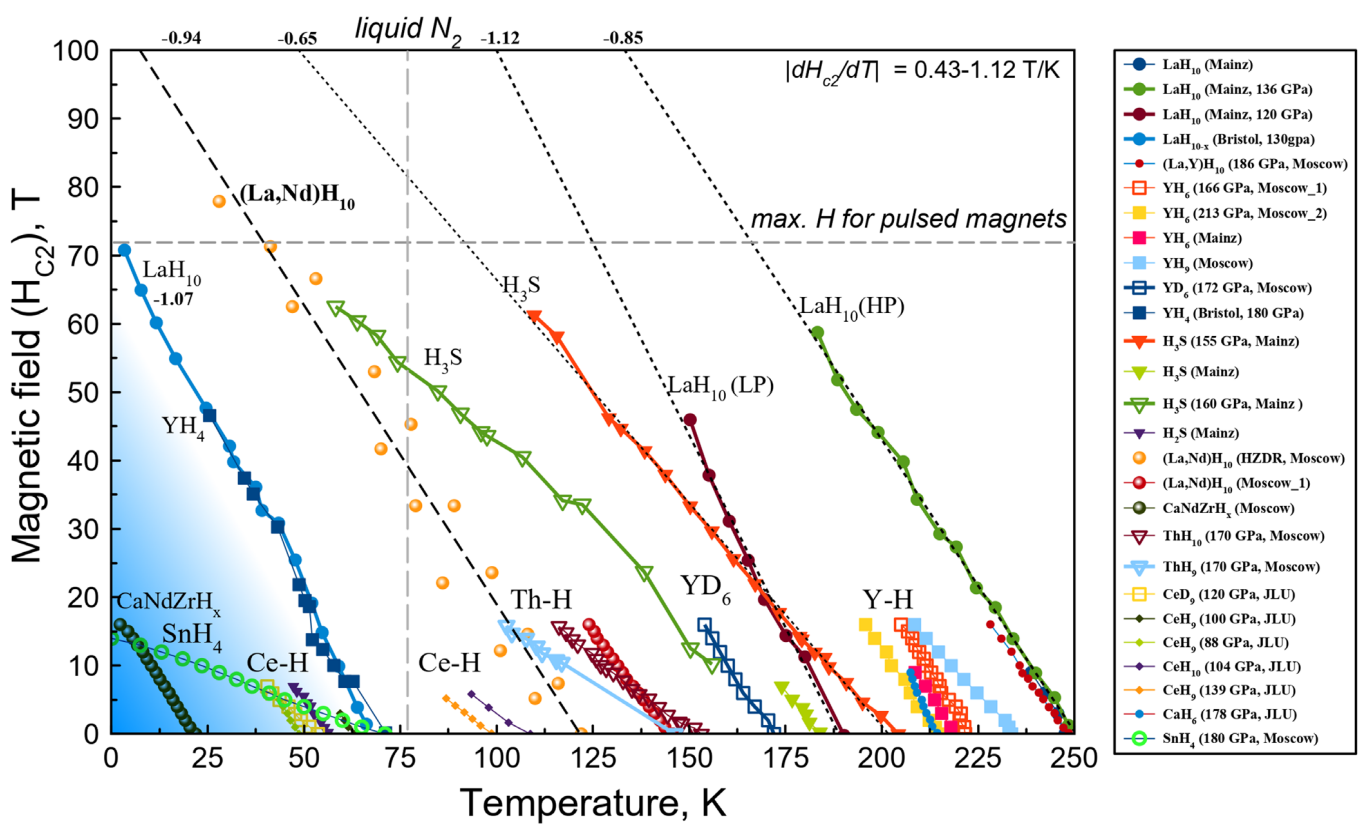

**Figure S18.** Currently known part of the magnetic phase diagram of superhydrides. The captions on the right show the proposed chemical formula and the scientific group that investigated the compound (HP indicates high-pressure synthesis, LP — low-pressure synthesis). As far as we know, no high-$T_C$ polyhydride has a well-defined saturation of $H_{C2}(T)$ at low temperatures predicted by the Werthamer–Helfand–Hohenberg (WHH) theory.

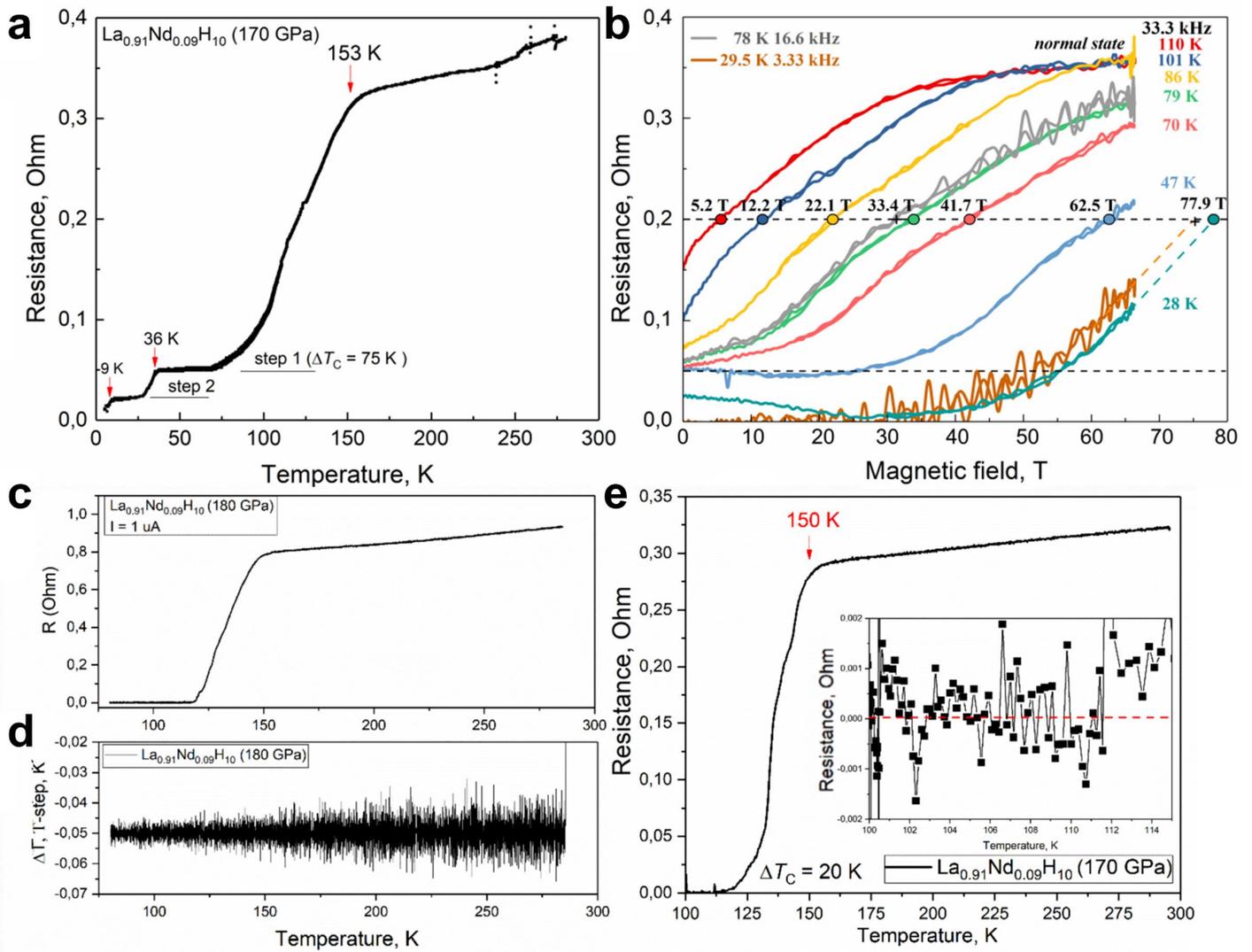

**Figure S19.** Electrical resistance of (La,Nd)$H_{10}$ with 9 at% of Nd at 170-180 GPa studied within the second series of high-field measurements (Nov. 2021, run 2). (a) Electrical resistance of the (La,Nd)$H_{10}$ sample (DAC E1) versus temperature after experiments in pulsed magnetic fields, warming cycle. (b) Dependence of the real part of the electrical resistance on the magnetic field. (c, d) The analysis of the cooling $T$ step for the sample in DAC E0, the importance of which for experiments in high-pressure cells was pointed out by J. Hirsch,[40,41] shows the absence of anomalies near $T_C$. (e) Superconducting transition and the initial temperature dependence of the $La_{0.91}Nd_{0.09}H_{10}$ resistance at 170 GPa in DAC E1 (before experiments in pulsed fields). This sample was used for pulsed magnetic measurements in run 2. Inset: residual resistance of the sample after the superconducting transition.

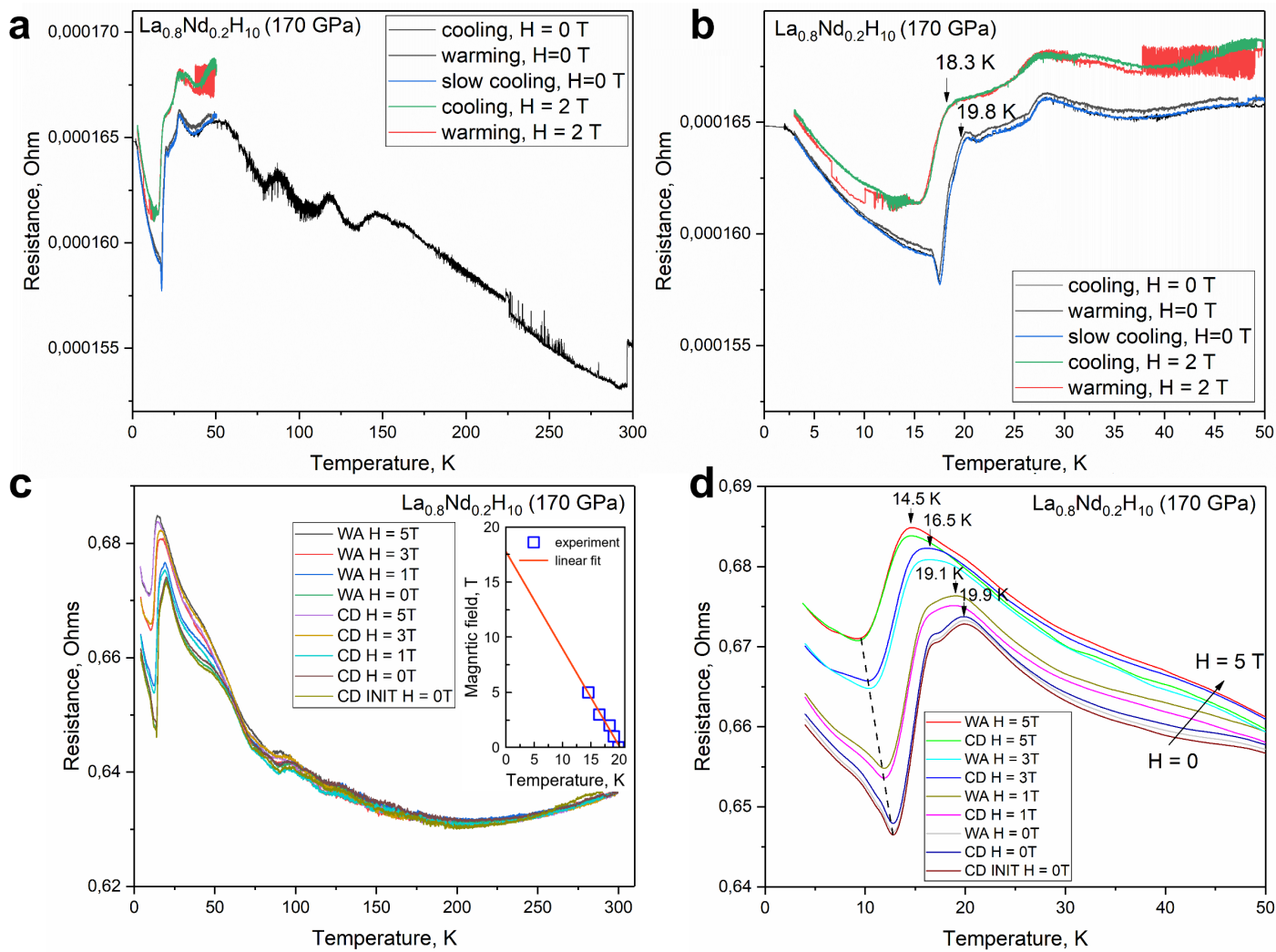

**Figure S20.** (a, b) Temperature and magnetic field (at $H = 0$ and 2 T) dependences of the resistance for $La_{0.8}Nd_{0.2}H_{10}$ sample at 170 GPa in DAC E2, cooling and warming cycles. $R(T)$ shows a nonmetallic character with signs of instability of the electrode system. (c, d) The same measurements for the $La_{0.8}Nd_{0.2}H_{10}$ sample performed after keeping the DAC E2 for 60 days. The character of $R(T)$ changed, but electrical resistance anomaly at ~20 K remains the same. This anomaly shifts to the low temperature region when an external magnetic field is applied.

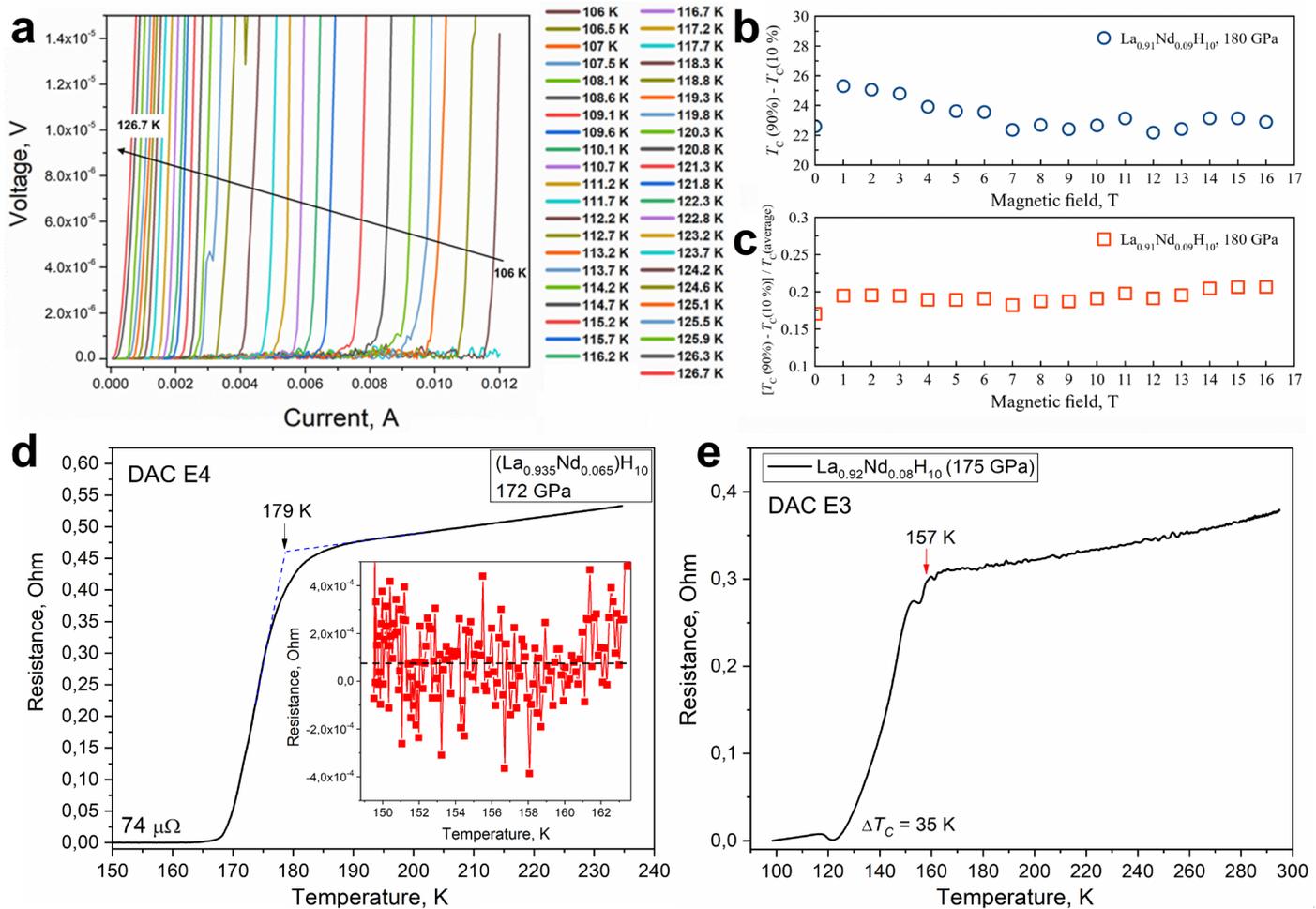

**Figure S21.** (a) Current–voltage characteristics of the DAC E0 at 180 GPa in the temperature range of 106–126.7 K in the absence of an external magnetic field. (b, c) Absolute and relative broadening of the superconducting transitions in (La,Nd)$H_{10}$ (DAC E1) in low magnetic fields of 0–16 T. (d) Temperature dependence of electrical resistance for the (La$_{0.935}$Nd$_{0.065}$)$H_{10}$ sample in DAC E4. Inset: residual resistance (averaged, 74 μOhm) of the sample below the superconducting transition temperature. (e) Temperature dependence of electrical resistance for the (La$_{0.92}$Nd$_{0.08}$)$H_{10}$ sample in DAC E3.

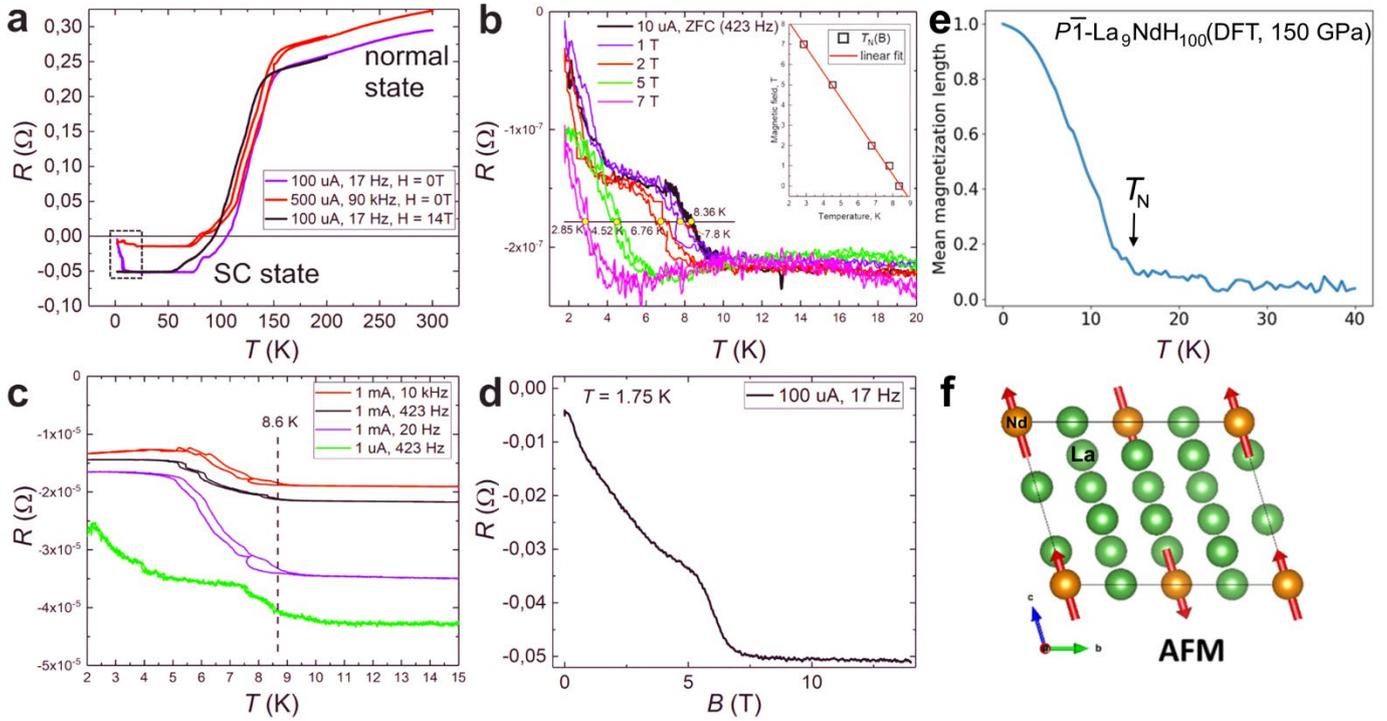

**Figure S22.** The low-temperature resistance anomaly in the (La,Nd)H$_{10}$ sample doped with 9 at% of Nd at 170 GPa (DAC E1). The formally negative resistance below 50 K is due to the reactive component of the impedance (parasitic capacitance) that occurs when a sinusoidal signal is used for measurements. (a) Wide superconducting transitions (75–150 K) and a jump in the electrical resistance in the low-temperature region (< 20 K), which is observed regardless of the frequency (17 Hz and 90 kHz) and current (0.1 and 0.5 mA) used. (b) Resistance $R(T)$ anomaly in various external magnetic fields up to 7 T. (c) The onset transition temperature about 8.6 K observed in the zero-field cooling and warming using different frequencies (20, 423 Hz, and 10 kHz) and excitation currents (1 μA and 1 mA). (d) Dependence of the sample resistance at 1.75 K on the applied magnetic field. The anomalous resistance behavior disappears at $H$ > 7.5 T. We cannot rule out that this anomaly is due to the presence of unknown superconducting impurity. (e) Calculated Néel temperature of destruction of the antiferromagnetic (AFM) ordering and (f) the most stable magnetic structure (AFM) of the $P\bar{1}$-La$_9$NdH$_{100}$ at 150 GPa.

# Calculations of the magnetic ordering of La–Nd hydrides

A possible interplay between superconductivity and magnetism in lanthanide polyhydrides is one of the most motivating factors for their investigation. Nd is the first metal in the lanthanide series whose hydrides should be magnetically ordered.[34] (La,Nd)H$_{10}$ studied in this work is possibly the first example of a polyhydride in which the magnetic ordering and the superconducting state coexist on the magnetic phase diagram and competitively suppress each other at ~9 K (Figure S22).

We explored the magnetic properties of $P\bar{1}$-La$_9$NdH$_{100}$, starting with determining its most stable collinear magnetic ordering - ferromagnetic (FM) or antiferromagnetic (AFM). Eight trial configurations were examined, one FM and seven AFM (Figure S23). The AFM configurations are exhaustive up to the supercell La$_{18}$Nd$_2$H$_{200}$ and were generated with the help of the derivative structure enumeration library enumlib.

The converged parameters of our calculations are: 670 eV for the kinetic energy cutoff of the plane wave basis set, $l = 30$ for the automatic generation of Γ-centered Monkhorst–Pack grids as implemented in the VASP code, and the smearing parameter $σ = 0.1$ with the Methfessel–Paxton method of order 1. These parameters give us a maximum error of 1 meV/atom with respect to more accurate calculations; they were also used in the study of the remaining phases $P4/mmm$-LaNdH$_{20}$, $C2/m$-La$_4$NdH$_{50}$, and $I4/m$-La$_4$NdH$_{50}$. We performed the full relaxation at 150 GPa without the spin–orbit coupling for the eight magnetic configurations. For some configurations, the magnetic moments changed orientation after the relaxation (Table S6).

The most stable collinear configuration for $P\bar{1}$-La$_9$NdH$_{100}$ is AFM2. The Ising Hamiltonian was used to model the magnetic interactions, considering them only up to first nearest neighbors. We used the Ising Hamiltonian together with the relaxed enthalpies of the FM and AFM2 configurations from Table S6 to compute the coupling constant $J_1$. The critical temperature of our Heisenberg model was obtained from a Monte Carlo simulation as implemented in VAMPIRE software. The size of the simulation box was 25×25×25 nm after the convergence tests, the transition temperature was taken as the value where the normalized mean magnetization length goes below 0.25. The obtained values of the coupling constant and Néel temperature (the point above which certain AFM material becomes paramagnetic) are shown in Table S7, the mean magnetization length plotted against the temperature — in Figure S24.

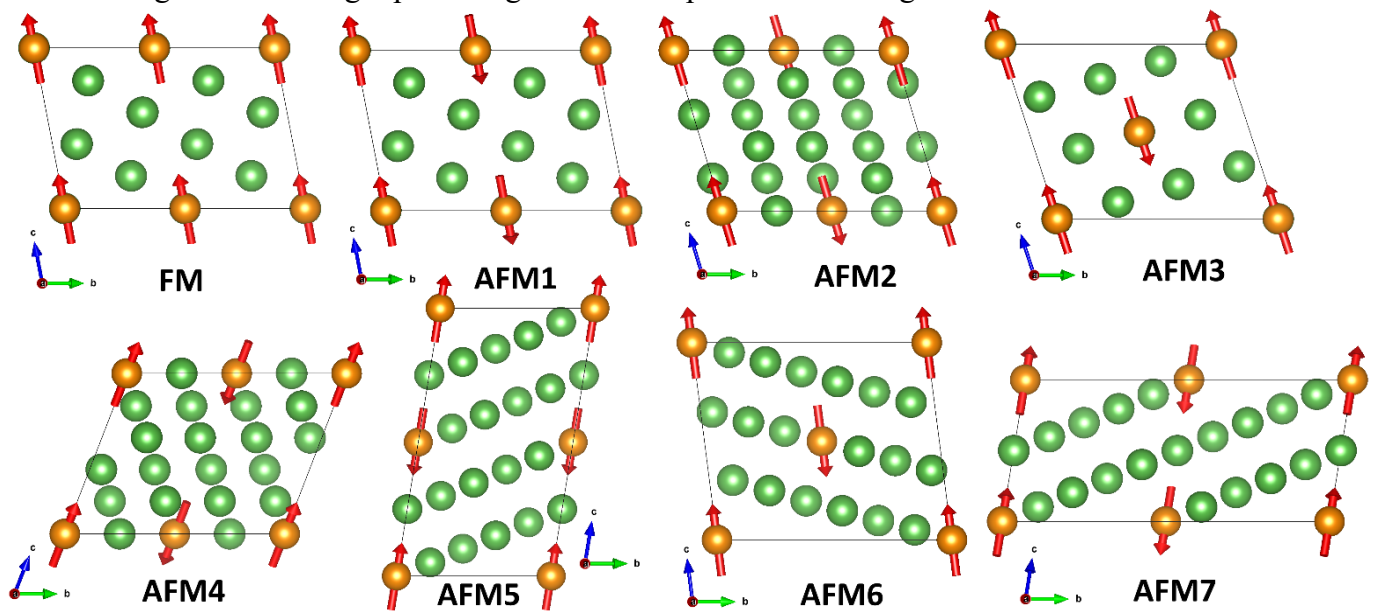

**Figure S23.** Eight trial magnetic configurations for $P\bar{1}$-La$_{18}$Nd$_2$H$_{200}$. For clarity, the hydrogen atoms are not shown.

**Table S6.** Enthalpy and magnetic ordering of the trial magnetic configurations for $P\bar{1}$-La$_9$NdH$_{100}$ after the full relaxation without the spin-orbit coupling.

| Initial configuration | Final configuration | Enthalpy, meV/Nd atom | Initial magnetic moments, $\mu_B$ | Final magnetic moments, $\mu_B$ |
|---|---|---|---|---|
| FM | FM | 406.9 | [+4, +4] | [+3.004, +1.012] |
| AFM1 | AFM1 | 4.36 | [+4, -4] | [+3.001, -3.001] |
| AFM2 | AFM2 | 0 | [+4, -4] | [+3.002, -3.002] |
| AFM3 | FM | 421.46 | [+4, -4] | [+3.002, +1.024] |
| AFM4 | AFM4 | 263.12 | [+4, -4] | [+3.303, -3.003] |
| AFM5 | AFM5 | 259.9 | [+4, -4] | [+3.003, -3.318] |
| AFM6 | - | 813.5 | [+4, -4] | [+3.003, -0.080] |
| AFM7 | AFM7 | 244.25 | [+4, -4] | [+2.999, -3.312] |

**Table S7.** Coupling constant $J_1$ and the estimated Néel temperature of $P\bar{1}$-La$_9$NdH$_{100}$.

| $J_1$, J/link | $T_N$, K |
|---|---|
| $-3.2596 \times 10^{-20}$ | 12 |

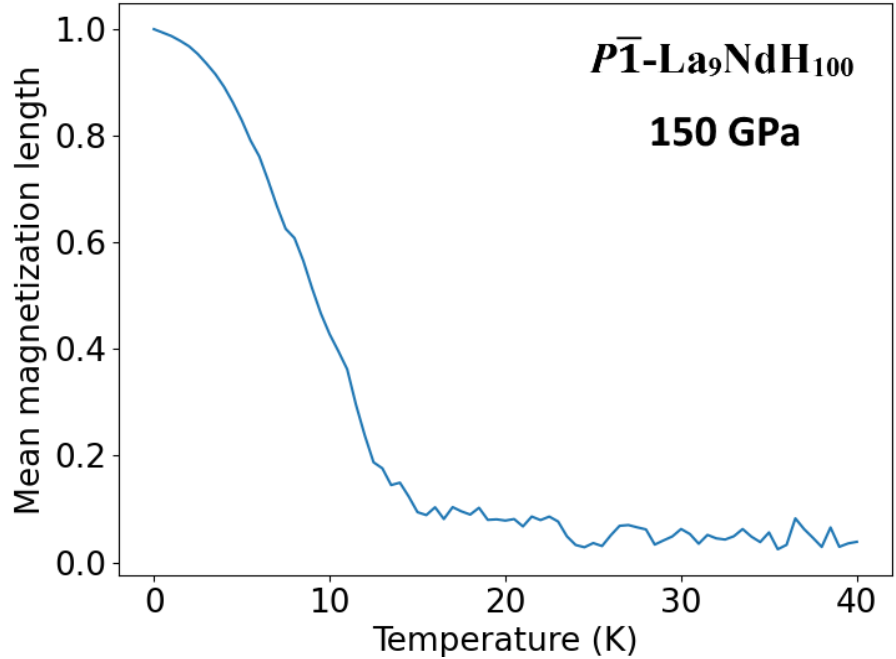

**Figure S24.** Normalized mean magnetization length of the spin-up channel with respect to the temperature in $P\bar{1}$-La$_9$NdH$_{100}$, obtained using the Monte Carlo simulation.

For $P4/mmm$-LaNdH$_{20}$, we also examined eight trial configurations, one FM and seven AFM (Figure S25), which are exhaustive up to the supercell La$_2$Nd$_2$H$_{40}$. We performed the full relaxation at 150 GPa without the spin–orbit coupling. For some configurations, the magnetic moments changed orientation after the relaxation (Table S8).

The most stable collinear configuration for $P4/mmm$-LaNdH$_{20}$ is FM. For this phase, we cannot estimate the critical temperature because the system of equations arising from the effective Hamiltonian is singular.

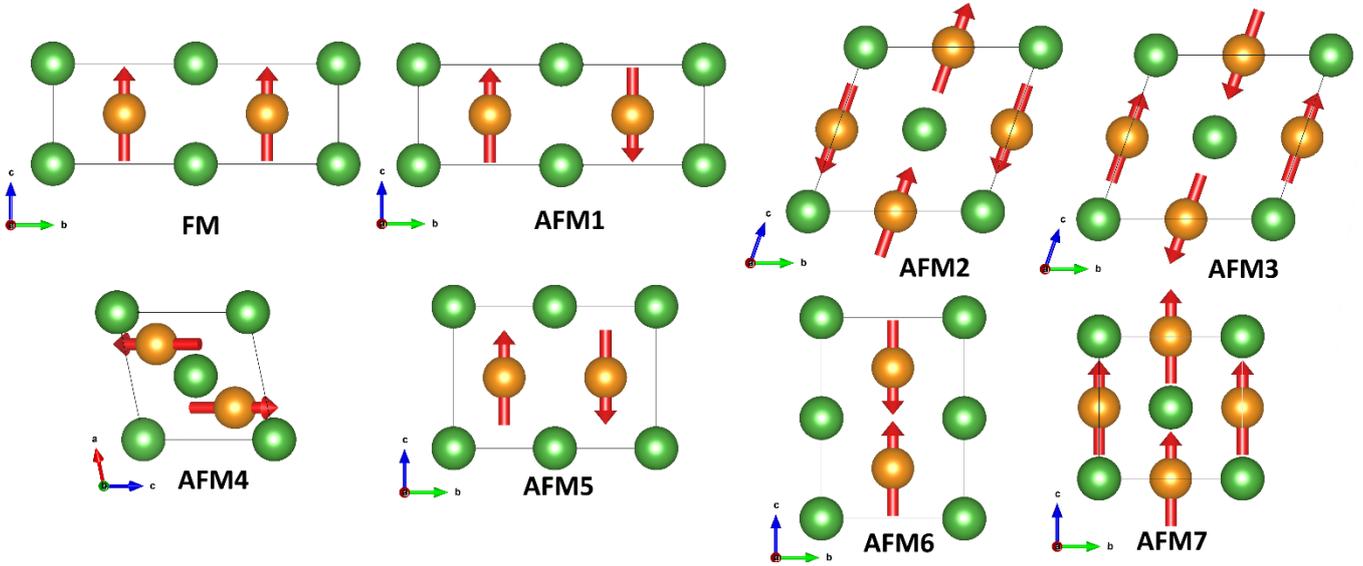

**Figure S25**. Trial magnetic configurations for $P4/mmm$-LaNdH$_{20}$. For clarity, the hydrogen atoms are not shown.

**Table S8.** Enthalpy and magnetic ordering of the trial magnetic configurations for $P4/mmm$-LaNdH$_{20}$ after the full relaxation without the spin–orbit coupling. NM stands for nonmagnetic.

| Initial configuration | Final configuration | Enthalpy, meV/Nd atom | Initial magnetic moments, $\mu_B$ | Final magnetic moments, $\mu_B$ |
|---|---|---|---|---|
| FM | FM | 0 | [+4, +4] | [+2.249, +2.249] |
| AFM1 | NM | 319.72 | [+4, -4] | [+0.035, -0.035] |
| AFM2 | AFM2 | 147.11 | [+4, -4] | [-2.077, +2.077] |
| AFM3 | AFM3 | 148.25 | [+4, -4] | [-2.077, +2.077] |
| AFM4 | NM | 325.65 | [+4, -4] | [-0.041, -0.038] |
| AFM5 | NM | 323.74 | [+4, -4] | [+0.083, -0.075] |
| AFM6 | NM | 323.31 | [+4, -4] | [+0.083, -0.075] |
| AFM7 | AFM7 | 107.83 | [+4, -4] | [-2.209, +2.209] |

Similarly, one FM and seven AFM trial configurations were also examined for $I4/m$-La$_4$NdH$_{50}$ (Figure S26) and $C2/m$-La$_4$NdH$_{50}$ (Figure S27). These configurations are exhaustive up to the supercell La$_8$NdH$_{100}$. The full relaxation was performed at 150 GPa without the spin–orbit coupling. After the relaxation, the magnetic moments changed the orientation for some configurations (Tables S9 and S10).

The most stable collinear configuration for $I4/m$-La$_4$NdH$_{50}$ is AFM2. The critical temperature cannot be estimated for this phase because the system of equations arising from the effective Hamiltonian is singular.

In the case of $C2/m$-La$_4$NdH$_{50}$, all configurations relaxed to the nonmagnetic state, so the collinear analysis shows that the material is nonmagnetic. We do not exclude the presence of nonzero magnetic moments after noncollinear calculations.

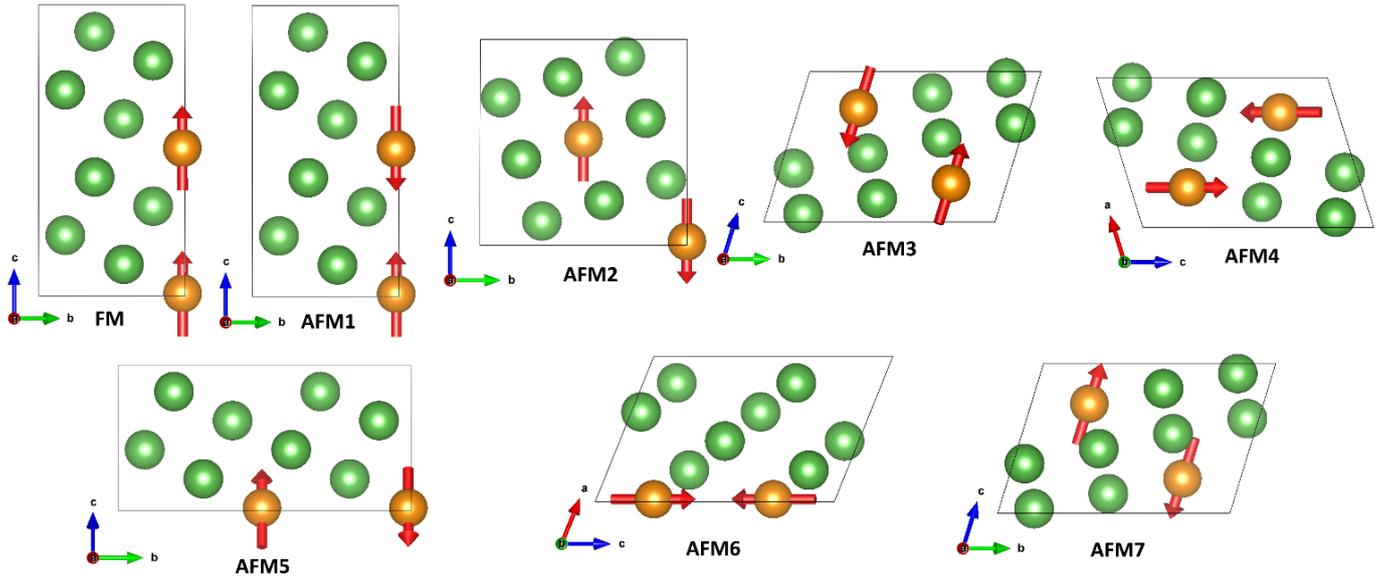

**Figure S26.** Trial magnetic configurations for $I4/m$-La$_4$NdH$_{50}$. For clarity, the hydrogen atoms are not shown.

**Table S9.** Enthalpy and magnetic ordering of the trial magnetic configurations for $I4/m$-La$_4$NdH$_{50}$ after the full relaxation without the spin–orbit coupling. NM stands for nonmagnetic.

| Initial configuration | Final configuration | Enthalpy, meV/Nd atom | Initial magnetic moments, $\mu_B$ | Final magnetic moments, $\mu_B$ |
|---|---|---|---|---|
| FM | NM | 137.74 | [+4, +4] | [+0.003, +0.003] |
| AFM1 | NM | 139.58 | [+4, -4] | [-0.003, +0.003] |
| AFM2 | AFM2 | 0.0 | [+4, -4] | [+1.879, -1.879] |
| AFM3 | AFM3 | 31.68 | [+4, -4] | [+1.876, -1.876] |
| AFM4 | AFM4 | 30.3 | [+4, -4] | [+1.871, -1.871] |
| AFM5 | NM | 130.62 | [+4, -4] | [+0.000, -0.000] |
| AFM6 | AFM6 | 36.63 | [+4, -4] | [+1.842, -1.842] |
| AFM7 | AFM7 | 44.82 | [+4, -4] | [+1.836, -1.836] |

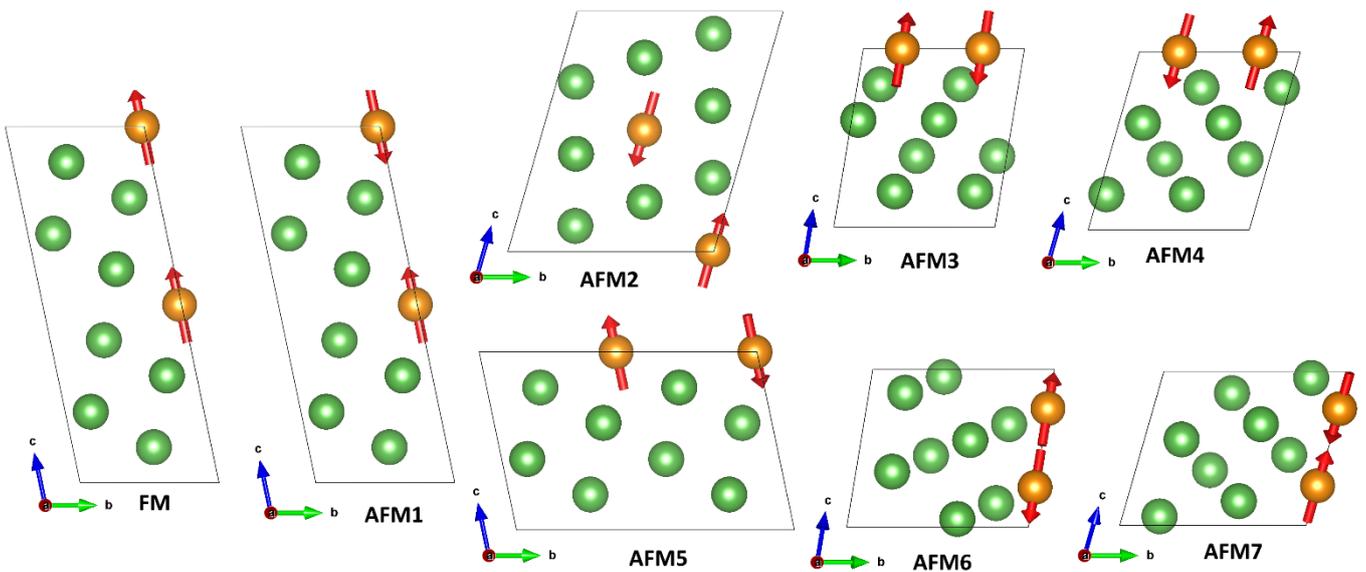

**Figure S27.** Trial magnetic configurations for $C2/m$-La$_4$NdH$_{50}$. For clarity, the hydrogen atoms are not shown.

**Table S10.** Enthalpy and magnetic ordering of the trial magnetic configurations for $C2/m$-La$_4$NdH$_{50}$ after the full relaxation without the spin–orbit coupling. NM stands for nonmagnetic, * means that the calculations did not converge after 300 ionic steps.

| Initial configuration | Final configuration | Enthalpy, meV/Nd atom | Initial magnetic moments, $\mu_B$ | Final magnetic moments, $\mu_B$ |
|---|---|---|---|---|
| FM | NM | 77.84 | [+4, +4] | [+0.000, +0.000] |
| AFM1 | - | 15.26 | [+4, -4] | [-0.069, -1.912] |
| AFM2 | NM | 91.33 | [+4, -4] | [-0.026, +0.026] |
| AFM3 | NM | 39.43 | [+4, -4] | [+0.037, +0.022] |
| AFM4 | NM | 0.03 | [+4, -4] | [-0.064, +0.031] |
| AFM5 | NM | 76.74 | [+4, -4] | [+0.000, -0.000] |
| AFM6 | * | * | [+4, -4] | * |
| AFM7 | NM | 0. | [+4, -4] | [-0.064, +0.031] |